\documentclass[prd,superscriptaddress,nofootinbib,twocolumn]{revtex4}
\usepackage{amsmath,amssymb,epsfig}

\usepackage{graphicx}
\usepackage{bm}

\usepackage{scalerel}
\usepackage{tikz}
\usetikzlibrary{svg.path}

\definecolor{orcidlogocol}{HTML}{A6CE39}
\tikzset{
	orcidlogo/.pic={
		\fill[orcidlogocol] svg{M256,128c0,70.7-57.3,128-128,128C57.3,256,0,198.7,0,128C0,57.3,57.3,0,128,0C198.7,0,256,57.3,256,128z};
		\fill[white] svg{M86.3,186.2H70.9V79.1h15.4v48.4V186.2z}
		svg{M108.9,79.1h41.6c39.6,0,57,28.3,57,53.6c0,27.5-21.5,53.6-56.8,53.6h-41.8V79.1z M124.3,172.4h24.5c34.9,0,42.9-26.5,42.9-39.7c0-21.5-13.7-39.7-43.7-39.7h-23.7V172.4z}
		svg{M88.7,56.8c0,5.5-4.5,10.1-10.1,10.1c-5.6,0-10.1-4.6-10.1-10.1c0-5.6,4.5-10.1,10.1-10.1C84.2,46.7,88.7,51.3,88.7,56.8z};
	}
}

\newcommand\orcidicon[1]{\href{https://orcid.org/#1}{\mbox{\scalerel*{
				\begin{tikzpicture}[yscale=-1,transform shape]
				\pic{orcidlogo};
				\end{tikzpicture}
			}{1}}}}

\usepackage{natbib,ifthen}
\usepackage{xcolor}
\usepackage[hidelinks]{hyperref}

\begin{document}

\newcommand{\fixme}[1]{{\textbf{Fixme: #1}}}
\newcommand{\detD}{{\det\!\cld}}
\newcommand{\clh}{\mathcal{H}}
\newcommand{\ud}{{\rm d}}
\renewcommand{\eprint}[1]{\href{http://arxiv.org/abs/#1}{#1}}
\newcommand{\adsurl}[1]{\href{#1}{ADS}}
\newcommand{\ISBN}[1]{\href{http://cosmologist.info/ISBN/#1}{ISBN: #1}}
\newcommand{\jcap}{J.\ Cosmol.\ Astropart.\ Phys.}
\newcommand{\mnras}{Mon.\ Not.\ R.\ Astron.\ Soc.}
\newcommand{\progress}{Rep.\ Prog.\ Phys.}
\newcommand{\prlett}{Phys.\ Rev.\ Lett.}
\newcommand{\procspie}{Proc.\ SPIE}
\newcommand{\na}{New Astronomy}
\newcommand{\apjl}{ApJ.\ Lett.}
\newcommand{\physrep}{Physics Reports}
\newcommand{\aap}{A\&A}
\newcommand{\aapr}{A\&A Rev.}

\newcommand{\ThreeJSymbol}[6]{\begin{pmatrix}
#1 & #3 & #5 \\
#2 & #4 & #6
\end{pmatrix}}

\newcommand{\SixJSymbol}[6]{\begin{Bmatrix}
#1 & #3 & #5 \\
#2 & #4 & #6
\end{Bmatrix}}

\defcitealias{2019MNRAS.486.5061R}{R19}
\defcitealias{2015MNRAS.450..317C}{C15}


\title[Conditional cosmological statistics]{The impact of our local environment on cosmological statistics}

\author{Alex Hall \orcidicon{0000-0002-3139-8651}}
\email{ahall@roe.ac.uk}
\affiliation{Institute for Astronomy, University of Edinburgh, Royal Observatory, Blackford Hill, Edinburgh, EH9 3HJ, U.K.}



\begin{abstract}
  We conduct a thorough investigation into the possibility that residing in an overdense region of the Universe may induce bias in measurements of the large-scale structure. We compute the conditional correlation function and angular power spectrum of density and lensing fluctuations while holding the local spherically averaged density fixed and show that for Gaussian fields this has no effect on the angular power at $l>0$. We identify a range of scales where a perturbative approach allows analytic progress to be made, and we compute leading-order conditional power spectra using an Edgeworth expansion and second-order perturbation theory. We find no evidence for any significant bias to cosmological power spectra from our local density contrast. We show that when smoothed over a large region around the observer, conditioning on the local density typically affects density power spectra by less than a percent at cosmological distances, below cosmic variance. We find that while typical corrections to the lensing angular power spectrum can be at the 10\% level on the largest angular scales and for source redshifts $z_s \lesssim 0.1$, for the typical redshifts targeted by upcoming wide imaging surveys the corrections are sub-percent and negligible, in contrast to previous claims in the literature. Using an estimate of the local spherically averaged density from a composite galaxy redshift catalogue we find that the corrections from conditioning on our own local density are below cosmic variance and subdominant to other non-linear effects. We discuss the potential implications of our results for cosmology and point out that a measurement of the local density contrast may be used as a consistency test of cosmological models.
\end{abstract}

\maketitle

\section{Introduction}
\label{sec:intro}

Current and upcoming surveys of large-scale structure such as Euclid\footnote{\url{https://sci.esa.int/web/euclid}}, LSST\footnote{\url{https://www.lsst.org/}}, and DESI\footnote{\url{https://www.desi.lbl.gov/}} aim to place percent-level constraints on the dark energy paradigm and make measurements of the summed mass of neutrino species~\cite{2011arXiv1110.3193L, 2018arXiv180901669T}. With the huge leap in statistical constraining power that these surveys represent, many previously negligible systematic effects must be mitigated to ensure unbiased and precise constraints on cosmological models.

Recently it was pointed out in Ref.~\cite{2019MNRAS.486.5061R} (hereafter~\citetalias{2019MNRAS.486.5061R}) that a potential source of bias arises from neglecting to account for the impact of our local environment on summary statistics measured in cosmological surveys. All astronomical observations are made from our privileged position within a region of the Universe (the Local Group) with an above-average density. Since the density field has long-range spatial correlations, the distribution of large-scale structure conditioned on our local density should differ from the unconditional distribution. In particular we might expect two-point statistics -- the most common summary statistics used to infer cosmological parameters -- to acquire a correction when conditioned on the local density. \citetalias{2019MNRAS.486.5061R} computed this correction assuming Gaussian fields, claiming percent-level effects on the lensing angular power spectrum over a wide range of angular multipoles, i.e. at a level potentially important for a Euclid-like survey.

In this work we will critically examine the suggestion that power spectra measured in cosmological surveys can be biased by our local density at a significant level, building upon the work of \citetalias{2019MNRAS.486.5061R}. If there is an effect then clearly it must be limited to the clustering of only the most nearby cosmic structure -- the correlation length of the density field, defined say as the radius at which the correlation function first crosses zero, is of order $100 \, h^{-1} \mathrm{Mpc}$ at $z=0$, whereas cosmological surveys typically measure large-scale structure at hundreds or thousands of comoving megaparsecs. In the case of density fluctuations there will be a small residual correction, quantified by the conditional angular power spectrum or conditional correlation function, which we compute in this work. In the case of gravitational lensing we might expect a more significant correction, since lensing inevitably picks up contributions from nearby structure, albeit suppressed by geometric factors. In any case, a full calculation is necessary to check that any residual biases can be safely ignored by future surveys.

We will focus on corrections to the power spectra from spatial correlations in the density field in the case of Gaussian and weakly non-Gaussian fields. We will compute the correlation function and angular power spectrum \emph{conditional} on a fixed spherically averaged density field around the observer. We are interested in cosmological scales where fluctuations in the matter density are small. It will transpire that in the case of purely Gaussian fields only the power at $l=0$ is affected, so for observable effects we need to go to at least second-order in the density field. Our approach is close to that suggested by \citetalias{2019MNRAS.486.5061R} where it was noted that the non-Gaussian calculation involves highly oscillatory multi-dimensional integrals. We will show how many of these integrals can be done analytically in the perturbative regime.

In the case of lensing, the local density field contributes to the unconditional lensing power at a statistical level through the matter power spectrum projected onto our past-light cone. Rather than including local fluctuations in the variance of the signal in this way, we study the impact of fixing the density in a local region to some measured value. Since the local density cannot be averaged when considering many lines of sight we might expect a conditional power spectrum to provide a better fit to observations. We will discuss the implications (if any) of conditioning on the local density on the information, bias, and consistency of cosmological models.

The effects considered in this work are distinct from other local effects which impact cosmological observables. For example, the local gravitational potential in which we reside adds a small blueshift to the spectrum of all extragalactic objects~\cite{2015JCAP...07..025W}. In addition, temporal variations of our local gravitational potential can add a Rees-Sciama/Integrated Sachs-Wolfe contribution to the cosmic microwave background (CMB)~\cite{2006MNRAS.369L..27R, 2007A&A...476...83M, 2010MNRAS.406...14F}. In this work we focus on the impact of fixing the local density contrast when computing ensemble averages of density and lensing anisotropies. The impact of the local density field on velocity statistics has been considered in Refs.~\cite{2014MNRAS.438.1805W, 2017MNRAS.467.2787H, 2018PhRvD..97j3519H}.

This paper is structured as follows. In Section~\ref{sec:cond_cls} we present the calculation of the conditional correlation function and angular power spectrum of density fluctuations and estimate the local spherically averaged density from a galaxy redshift catalogue. The main results of this section are Equation~\eqref{eq:cond_cov_tree} and Equation~\eqref{eq:bll0_s}. In Section~\ref{sec:lensing} we compute the conditional lensing angular power spectrum, the main result being Equation~\eqref{eq:limber}. In Section~\ref{sec:discussion} we discuss the implications of our results for cosmology and discuss the use of the local density as a consistency check on cosmological models. We conclude in Section~\ref{sec:conclusions}.

For numerical work we assume a flat $\Lambda$CDM cosmological model with parameters fixed to the best-fit values from Planck 2015 (TT, TE, EE + lowP + lensing + ext)~\cite{2016A&A...594A..13P}, i.e. $(\Omega_b h^2, \Omega_ch^2, h, A_s, n_s) = (0.0223, 0.1188, 0.6774, 2.142 \times 10^{-9}, 0.9667)$. We will set $c=1$ unless otherwise specified.

\section{Conditional cosmological statistics}
\label{sec:cond_cls}

The goal of this section is to compute the angular power spectrum and correlation function of the matter density contrast conditioned on the local density fluctuation in which we reside. We define the local density contrast $\delta_0(R)$ to be the density field $\delta(\mathbf{r})$ smoothed with a spherical top-hat filter of comoving radius $R$ and located at the origin, i.e.
\begin{equation}
  \delta_0(R) \equiv \frac{3}{4\pi R^3} \int \mathrm{d}^3\mathbf{r} \, \Theta(R - |\mathbf{r}|) \delta(\mathbf{r}),
\end{equation}
where $\Theta$ is the Heaviside step function. The constrained random field whose angular power spectrum we seek to compute consists of realizations of $\delta(\mathbf{r})$ which give rise to a fixed $\delta_0(R)$. We keep the time dependence of the density field implicit in this section and focus solely on the real-space dark matter density field.

\subsection{Order-of-magnitude estimate}
\label{subsec:order}

Before presenting a detailed calculation we first provide a rough order-of-magnitude estimate of the size of the correction to the angular power spectrum that results from conditioning on our local density.

Firstly, suppose $\delta(\mathbf{r})$ obeyed Gaussian statistics with mean zero and covariance $\xi(d) \equiv \langle \delta(\mathbf{r}_1) \delta(\mathbf{r}_2) \rangle$ where $d \equiv |\mathbf{r}_2 - \mathbf{r}_1|$ and we assume that the field is statistically homogeneous and isotropic. Angle brackets here denote the ensemble mean over the unconditional distribution of $\delta(\mathbf{r})$. Since smoothing is a linear operation the field given by $[\delta_0(R), \delta(\mathbf{r})]$ is also Gaussian, with mean zero\footnote{We will assume that $\mathbf{r} \neq 0$ to ensure the covariance matrix is invertible.}. Manipulating the Gaussian probability distribution for this field and using that $p(A | B) = p(A, B)/p(B)$ it is easy to show that the conditional distribution of $\delta(\mathbf{r})$ given $\delta_0(R)$ is also Gaussian, with mean $\langle \delta(\mathbf{r} )| \delta_0(R) \rangle = \langle \delta(\mathbf{r}) \delta_0(R) \rangle \delta_0(R)/\sigma^2(R)$, where $\sigma^2(R)$ is the variance of $\delta_0(R)$. The covariance of the conditional field is independent of the value of $\delta_0(R)$ and is given by $\mathrm{cov}[\delta(\mathbf{r}_1), \delta(\mathbf{r}_2) | \delta_0(R)] = \xi(d) - \xi_R(r_{1})\xi_R(r_2)/\sigma^2(R)$ where $\xi_R(r_1) \equiv \langle \delta(\mathbf{r}_1)\delta_0(R) \rangle$. When $\mathbf{r}_1 = \mathbf{r}_2$ the correction to the variance at a point is negative, representing a loss of variance due to part of the field being held fixed. As expected, when $r_1$ and $r_2$ are large the correction becomes negligible since the field decorrelates from the local density. Crucially for what follows, in the Gaussian case the correction is independent of the angle between the two points $\mathbf{r}_1$ and $\mathbf{r}_2$ and depends only on their radial distances from the observer. In spherical harmonic space this corresponds to a correction to the (unobservable) $l=0$ mode. Therefore, interesting effects on the angular power spectrum can only arise at second-order where the density field is non-Gaussian. In simple terms, angle-dependent effects can only arise when $\delta(\mathbf{r}_1)$ and  $\delta(\mathbf{r}_2)$ couple with $\delta(\mathbf{r}_0)$, and this can only happen in the presence of a connected three-point function, which Gaussian fields do not possess. Although the independence of the correction from the angular separation in the Gaussian case was noted in \citetalias{2019MNRAS.486.5061R}, they claim an effect at all $l$, contrary to the above argument.

Thus, corrections to the $l>0$ angular power spectrum only arise at non-linear order in the density field. An estimate of the size of this correction can be made using techniques similar to those employed in the study of non-linear effects on the baryon acoustic oscillation peak. Since corrections from conditioning on the local density will be greatest when the two points considered are at least within a correlation length $r_c$ of the observer, we consider the case $r_1, r_2 \ll r_c$. Demanding that the points are well-within a smoothing volume centered on the observer simplifies the following discussion, so we will additionally impose that $r_1, r_2 \ll R$. As shown in Ref.~\cite{2011JCAP...10..031B} and as follows from Birkhoff's theorem, short-wavelength density fluctuations within a spherically symmetric long-wavelength adiabatic overdensity such as $\delta_0(R)$ behave as if they reside in an FRW universe with slight positive curvature to first order in $\delta_0(R)$. As described in Refs.~\cite{2012PhRvD..85j3523S, 2014JCAP...05..048C}, this enhances the growth of short-wavelength fluctuations, rescales the background density, and dilates all distances within the long-wavelength fluctuation, resulting in a correction to the correlation function in Einstein-de Sitter of $ \left[68\xi(d)/21 + d \xi'(d)/3 \right]\delta_0(R)$ to first order in $\delta_0(R)$, where a prime denotes differentiation. For a power-law correlation function $\xi(d) \sim d^{n}$ this implies a fractional correction of $(68/21 + n/3)\delta_0(R)$. Thus, except in cases of extremely negative $n$, conditioning on a positive local density fluctuation \emph{enhances} the correlation of fluctuations well within a smoothing volume. In the case of gravitational lensing, we expect a similar enhancement for the nearest source redshifts and on the largest angular scales.

Since the above expressions hold only at first order in $\delta_0(R)$, we must choose $R$ to be much greater than the scale of non-linearity. Then $\delta_0(R) \ll 1$ and hence the fractional correction from conditioning is expected to be very small. We can quantify this further by estimating our own local $\delta_0(R)$ from galaxy redshift surveys. For this purpose we use the 2M++ galaxy redshift catalogue~\cite{2011MNRAS.416.2840L}, a composite of the 2MASS Redshift Survey (2MRS), SDSS-DR7, and 6dF with depth $K_{\mathrm{2M++}} \leq 11.5$ over the full sky (excluding the galactic plane), increasing to $K_{\mathrm{2M++}} \leq 12.5$ in regions covered by SDSS and 6dF. The survey has full-sky coverage out to comoving distance $125 \, h^{-1}\mathrm{Mpc}$, corresponding to the magnitude limit of 2MRS. This sample has been used extensively to test for local voids which might bias the interpretation of the locally measured Hubble rate~\cite{2011ApJ...730..119R, 2017MNRAS.471.4946W, 2019ApJ...875..145K}. We use the publicly available luminosity-weighted galaxy density field from Ref.~\cite{2015MNRAS.450..317C}\footnote{\url{https://cosmicflows.iap.fr/}} (hereafter \citetalias{2015MNRAS.450..317C}), which is weighted to account for incompleteness and normalized to a constant effective luminosity-weighted galaxy bias $b^*$. Using the measurement of $f/b^*$ from \citetalias{2015MNRAS.450..317C} and our cosmology we find that $b^* = 1.23 \pm 0.06$ for the 2M++ sample. We convert the galaxy density into a dark matter density with this bias and then average in spheres around the observer.

\begin{figure}
  \includegraphics[width=\columnwidth]{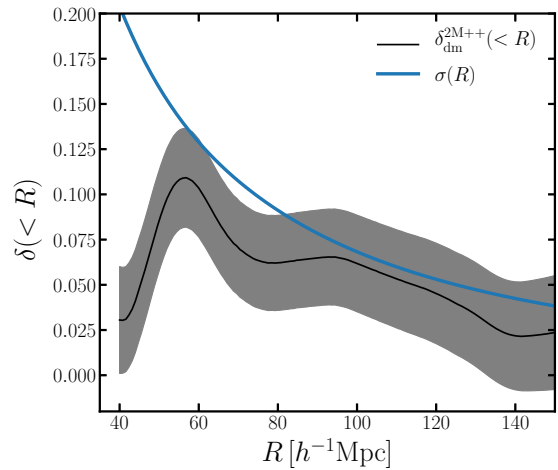}
  \caption{The local dark matter density contrast averaged in spheres of radius $R$ around the observer (solid black), as measured from the 2M++ galaxy density field of \citetalias{2015MNRAS.450..317C}, and assuming a constant luminosity-weighted galaxy bias of $b^* = 1.23 \pm 0.06$ in $\Lambda$CDM. The grey bands represent the 1$\sigma$ error on this average from the combination of shot noise, uncertainty in the galaxy bias, and uncertainty in the global mean density. Note that these errors are highly correlated for neighbouring $R$ values. The solid blue curve is the spherically averaged standard deviation of the linear dark matter density field at $z=0$.}
  \label{fig:2Mpp_density}
\end{figure}

In Figure~\ref{fig:2Mpp_density} we plot the spherically averaged dark matter density contrast as a function of the smoothing radius $R$, as well as the linear standard deviation of the matter density field at $z=0$ in our cosmology, computed with \textsc{CAMB}~\cite{2000ApJ...538..473L}. We estimate the shot noise on the smoothed density field using the Schechter function fit to the luminosity function from Ref.~\cite{2011MNRAS.416.2840L} to estimate the number of galaxies in each pixel and add this in quadrature with the galaxy bias uncertainty. We then include the additional uncertainty incurred when estimating the global mean density from the average density in the 2M++ volume. Assuming the difference between the survey and global mean densities is small (justified since the linear r.m.s. density contrast on the survey scale is about 0.024), it is straightforward to show that this mismatch biases the density contrast on average by a multiplicative factor of $\alpha(R) = 1 - \xi(R,R_B)/\sigma^2(R)$, where $\xi(R,R_B)$ is the correlation between density contrasts averaged in spheres of radius $R$ and $R_B$, with $R_B \approx 200 \, h^{-1}\mathrm{Mpc}$ the survey scale. We account for this bias by dividing the dark matter density from \citetalias{2015MNRAS.450..317C} by $\alpha(R)$. The extra variance from super-survey fluctuations is then roughly $[\sigma^2(R_B) - \xi(R, R_B)^2/\sigma^2(R)]/\alpha(R)^2$, which we add in quadrature to the shot noise and bias uncertainty. The total error on $\delta_0(R)$ estimated this way is shown as the bands in Figure~\ref{fig:2Mpp_density}. Note that this procedure does not correctly account for the galaxy weights or the pixel smoothing kernel, so should be taken as a rough estimate of the noise. From Figure~\ref{fig:2Mpp_density} we see that the local matter density contrast smoothed on scales $R \sim 10^2 \, h^{-1}\mathrm{Mpc}$ (within the 2MRS redshift coverage) is of order $10^{-2}$. We can thus anticipate that corrections to cosmological two-point functions from conditioning on this local density will be at the percent level at most.

In Appendix~\ref{app:borg} we consider an alternative measurement of the local dark matter density field using the Bayesian reconstruction method of Ref.~\cite{2019A&A...625A..64J}. The spherically averaged density contrasts of the two methods are in agreement at the 2$\sigma$ level, but the Bayesian mean density is typically smaller and consistent with zero, i.e. consistent with zero correction to the power spectrum from conditioning. The results presented in this work are in units of the fluctuation $\delta_0(R)/\sigma(R)$, such that any estimate of $\delta_0(R)$ may be substituted to compute conditional statistics. For example if one believed that we reside within a large-scale void on some scale $R$ then one could compute conditional power spectra by substituting the appropriate $\delta_0(R)$.

We will soon see that the full calculation of the correction to the $l>0$ angular power spectrum is simplified considerably when the smoothing radius is chosen to ensure that $\delta_0(R)$ is linear. We will largely focus on the scale $R = 120 \, h^{-1}\mathrm{Mpc}$, which is sufficiently large to ensure linearity but sufficiently small that a reliable full-sky spherical average can be obtained from 2M++. Using the 2M++ density field from \citetalias{2015MNRAS.450..317C} described above, we find $\delta_0(R = 120 \, h^{-1}\mathrm{Mpc}) \approx 0.045 \pm 0.028$, corresponding to a fluctuation $\nu(R) \equiv \delta_0(R)/\sigma(R)$ of $\nu(R = 120 \, h^{-1}\mathrm{Mpc}) \approx 0.85 \pm 0.53$. We emphasise that the errors here are only rough estimates but are sufficiently accurate for our purposes. When measured with the BORG method in Appendix~\ref{app:borg} we find a smaller local density $\delta_0(R) = -0.024 \pm 0.042$, and hence a local fluctuation of $\nu(R) = -0.45 \pm 0.79$.

%
\begin{figure}
  \includegraphics[width=\columnwidth]{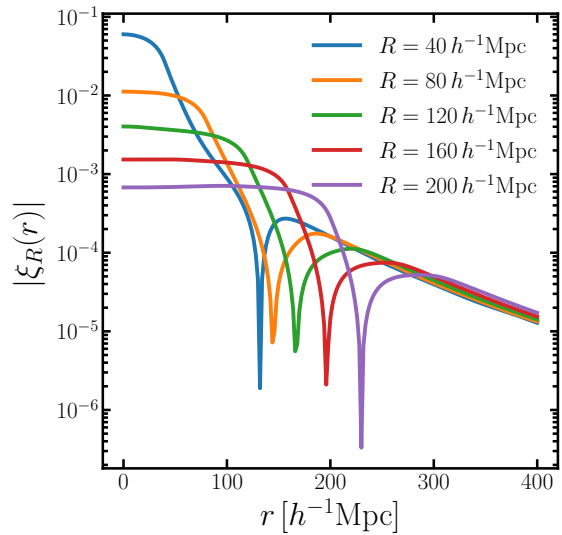}
  \caption{Absolute value of the correlation between the local density contrast averaged in spheres of radius $R$ with the unsmoothed density contrast at radius $r$. Curves are ordered top to bottom at $r=0$ for increasing $R$ and all go negative at large $r$. This function roughly dictates the radial dependence of corrections to the power spectrum from conditioning on the local density.}
  \label{fig:xi_Rs}
\end{figure}

In summary, we expect corrections to the angular power spectrum of density or lensing fluctuations for $l>0$ to be at most at the percent level, with the largest corrections for lensing coming from nearby source redshifts and large angular scales. We now turn to a detailed calculation of conditional angular power spectra. The radial dependence of the correction is expected to be primarily determined by the correlation function $\xi_R(r)$, which is plotted in Figure~\ref{fig:xi_Rs}. This figure shows that corrections should be approximately constant with distance for $r \lesssim R$ and rapidly dying away on scales greater than the local smoothing radius. As expected, for low $r$ the amplitude of the correction also rapidly decreases with increasing $R$, i.e. as the smoothing procedure becomes more aggressive. In the case of lensing fluctuations this radial dependence will be modified by a kernel dictated by the lensing geometry.

\subsection{Conditional Edgeworth expansion}
\label{subsec:edge}

To derive the conditional angular power spectrum of density and lensing fluctuations, we will first consider the more general problem of deriving conditional probability distributions. In the Gaussian case considered in Section~\ref{subsec:order} this was straightforward and amounted to manipulating Gaussian probability distributions and using the fundamental relation $p(A | B) = p(A, B)/p(B)$. We have seen however that observable effects can only arise at non-linear order where the field is non-Gaussian. In the non-Gaussian setting the calculation becomes more difficult since no fully general non-Gaussian distribution for the density field is known, and even if it were then extracting an analytic prediction for the conditional power spectrum seems intractable.

However, since the correction to the power spectrum from conditioning on the local density field is expected to be small for all practical purposes on cosmological scales, we may assume that the non-Gaussianity in $\delta(\mathbf{x})$ is itself small and use perturbative approaches to compute the leading order effects. The Edgeworth expansion of the probability density~\cite{1995ApJ...442...39J} then provides a useful approximation in the limit of weak non-Gaussianity which can be used to construct conditional statistics. The general approach we will take is, schematically, to use the leading order non-Gaussian forms of $p(A,B)$ and $p(B)$ from their Edgeworth expansions to construct an Edgeworth expansion for $p(A|B)$, from which we can read off the first few conditional cumulants. Note that this procedure does not suffer from the Edgeworth expansion's well-known problems in producing a positive-definite normalizable probability distribution - we only need the first few cumulants rather than the full distribution.

The calculation of the conditional Edgeworth expansion is presented in Ref.~\cite{10.2307/2336555, McCullagh} for a few simplifying cases. Here we present it in full generality, before specialising to the cosmological context. We follow the method and notation of Ref.~\cite{McCullagh}. Implicit summation over repeated indices should be assumed throughout unless specified otherwise.

The cumulants of a set of $p$ variables $X^{\alpha}$ (with $\alpha$ and all other Greek letters running from 1 to $p$) are denoted $\kappa^{\alpha}$, $\kappa^{\alpha, \beta}$, $\kappa^{\alpha, \beta, \gamma}$ etc. The probability density of $X^{\alpha}$ is denoted $f_X(x;\kappa)$, where the dependence on the set of cumulants is made explicit. Since we will ultimately divide a joint density by a marginalised density, it will prove more straightforward to work with log-densities where this division becomes a subtraction. $X^{\alpha}$ has an Edgeworth expansion around a Gaussian density given by
\begin{equation}
  \log f_X(x,\kappa) = \log \phi(x; \kappa) + \kappa^{\alpha, \beta, \gamma}h_{\alpha \beta \gamma}(x)/3! + ...,
  \label{eq:edge}
\end{equation}
where the Gaussian density $\phi(x;\kappa)$ is given by
\begin{equation}
  \phi(x;\kappa) \equiv \lvert 2\pi \kappa^{\alpha,\beta} \rvert^{-1/2} \exp \left[ -\frac{1}{2}(x^\alpha - \kappa^\alpha)(x^\beta - \kappa^\beta)\kappa_{\alpha, \beta}\right],
\end{equation}
where $\kappa_{\alpha, \beta}$ are elements of the matrix inverse of the covariance matrix $\kappa^{\alpha, \beta}$, and $\lvert \kappa^{\alpha,\beta} \rvert$ is the determinant of the covariance matrix. Also appearing in Equation~\eqref{eq:edge} is the Hermite tensor $h_{\alpha \beta \gamma}(x)$ given by
\begin{equation}
  h_{\alpha \beta \gamma}(x) = h_{\alpha}(x)h_{\beta}(x)h_{\gamma}(x) - h_{\alpha}(x)\kappa_{\beta,\gamma}[3],
\end{equation}
where the notation $[3]$ denotes permutations of the given partitioning of indices in the preceding expression (in the case above 3 terms result from this). The Hermite tensor $h_{\alpha}(x)$ is given by
\begin{equation}
  h_{\alpha}(x) = \kappa_{\alpha, \beta}(x^\beta - \kappa^\beta),
\end{equation}
i.e. the deviation from the mean normalized by the inverse covariance.

We assume that the non-Gaussianity of $X^{\alpha}$ is weak, such that the third cumulant $\kappa^{\alpha,\beta,\gamma}$ is one order higher in perturbation theory than the second cumulant, and higher-order cumulants are successively smaller. Then we can truncate the Edgeworth series by neglecting the higher-order terms denoted by the ellipsis in Equation~\eqref{eq:edge}.

Now, let us partition $X$ into two sets $X^{(1)}$ and $X^{(2)}$ of length $q$ and $p-q$ respectively. Elements of $X^{(1)}$ will be indexed by ${i,j,k,...}$ and elements of $X^{(2)}$ by ${r,s,t,...}$. We can decompose all the summations in Equation~\eqref{eq:edge} into these two blocks. For example, the Gaussian term becomes
\begin{align}
  & \log \phi(x;\kappa) = -\frac{1}{2}(x^i - \kappa^i)(x^j - \kappa^j)\kappa_{i,j} \nonumber\\
  &- \frac{1}{2}(x^i - \kappa^i)(x^r - \kappa^r)\kappa_{i,r}[2] - \frac{1}{2}(x^r - \kappa^r)(x^s - \kappa^s)\kappa_{r,s} \nonumber \\
  & - \frac{1}{2}\log \lvert \kappa^{i,j} - \kappa^{i,r}\bar{\kappa}_{r,s}\kappa^{s,j} \rvert - \frac{1}{2}\lvert \kappa^{r,s} \rvert \nonumber\\
  & - \frac{q}{2}\log 2\pi - \frac{(p-q)}{2}\log 2\pi,
\end{align}
where we used the determinant theorem for block matrices. Note that we have to be careful to distinguish $\kappa_{r,s}$ -- the $(r,s)$ block of the full inverse joint covariance matrix -- from $\bar{\kappa}_{r,s}$, hereafter defined as the matrix inverse of the $(r,s)$ block of the full covariance matrix $\kappa^{\alpha, \beta}$. These are related via the formulae for block matrix inversion by
\begin{equation}
  \kappa_{r,s} = \bar{\kappa}_{r,s} + \bar{\kappa}_{r,t} \, \kappa^{t,i}  \, \bar{\kappa}_{i,j}^{Sc}  \, \kappa^{j,u}  \, \bar{\kappa}_{u,s}
  \label{eq:wood}
\end{equation}
where $\bar{\kappa}_{i,j}^{Sc}$ are elements of the inverse of the Schur complement matrix whose elements are given by
\begin{equation}
  \kappa_{Sc}^{i,j} \equiv \kappa^{i,j} - \kappa^{i,r}\bar{\kappa}_{r,s}\kappa^{s,j}.
\end{equation}

We seek the density of $X^{(1)}$ conditional on some realization of $X^{(2)}$ - denote this realization by $x^r$. This density has an Edgeworth expansion that can be obtained by dividing the full joint density by the marginal density of $X^{(2)}$. This marginal density has an Edgeworth expansion given by
\begin{equation}
  \log f_{X^{(2)}}(x;\kappa) = \log \phi_2(x; \kappa) + \kappa^{r,s,t} \bar{h}_{rst}(x)/3! + ...,
\end{equation}
where the Gaussian part is given by
\begin{equation}
  \phi_2(x; \kappa) = \lvert 2\pi \kappa^{r,s} \rvert^{-1/2} \exp \left [-\frac{1}{2} (x^r - \kappa^r)(x^s - \kappa^s)\bar{\kappa}_{r,s} \right],
  \label{eq:marg_edge}
\end{equation}
and the Hermite tensor $\bar{h}_{rst}$ is given by
\begin{align}
  \bar{h}_{rst}(x) &= \bar{h}_{r}(x)\bar{h}_{s}(x)\bar{h}_{t}(x) - \bar{h}_{r}(x)\bar{\kappa}_{s,t}[3], \nonumber \\
  \bar{h}_r(x) &= \bar{\kappa}_{r,s}(x^s - \kappa^s).
\end{align}
Note that a key property here is that the marginal cumulants of $X^{(2)}$ are just given by the appropriate sub-block of the full joint cumulants $X^{\alpha}$. That this is true may be formally proved with the cumulant generating function. 
  
To get the (log) conditional density, we simply subtract Equation~\eqref{eq:marg_edge} from Equation~\eqref{eq:edge}, being careful to remember that sub-blocks of matrix inverses aren't the same as inverses of matrix sub-blocks, and that the two are related by the block-matrix inversion/Woodbury formulae Equation~\eqref{eq:wood}.

The derivation is laborious in the general case, so we only quote the results here. We find the conditional density as an Edgeworth expansion whose first three cumulants (i.e. conditional mean, covariance matrix, and three-point function) are given (a tilde on a quantity will denote a conditional quantity throughout) by
\begin{align}
  \tilde{\kappa}^i &= \kappa^i + \kappa^{i,r}\bar{h}_r + \kappa_c^{i,r,s}\bar{h}_{rs}/2, \nonumber \\
  \tilde{\kappa}^{i,j} &= \kappa^{i,j} - \kappa^{i,r} \bar{\kappa}_{r,s} \kappa^{s,j} + \kappa_c^{i,j,r}\bar{h}_r, \nonumber \\
  \tilde{\kappa}^{i,j,k} &= \kappa_c^{i,j,k},
  \label{eq:cond_cums}
\end{align}
where we have defined the quantity $\beta^i_r \equiv \kappa^{i,s}\bar{\kappa}_{s,r}$ and the Hermite tensor $\bar{h}_{rs}$ is given by $\bar{h}_{rs} = \bar{h}_r \bar{h}_s - \bar{\kappa}_{r,s}$. The conditional third cumulants in Equation~\eqref{eq:cond_cums} are given by
\begin{align}
  \kappa_c^{i,j,k} &= \kappa^{i,j,k} - \beta^i_r\kappa^{r,j,k}[3] + \beta^i_r \beta^j_s \kappa^{r,s,k}[3] - \beta^i_r \beta^j_s \beta^k_t \kappa^{r,s,t}, \nonumber \\
  \kappa_c^{i,j,r} &= \kappa^{i,j,r} - \beta^j_s\kappa^{i,r,s}[2] + \beta^i_t \beta^j_s \kappa^{r,s,t} , \nonumber \\
  \kappa_c^{i,r,s} &= \kappa^{i,r,s} - \beta^i_t\kappa^{t,r,s}.
  \label{eq:cond_3pt}
\end{align}
These expressions agree with those in Ref.~\cite{McCullagh} when $\kappa^{i,r} = 0$, and with those in Ref.~\cite{10.2307/2336555} when $\kappa^\alpha = 0$. Note that the conditional third cumulants are just the unconditional third cumulants of the decorrelated variables $(Y^{(1)},Y^{(2)})$ where $Y^i = X^i - \beta^i_rX^r$ and $Y^r = X^r$.

In the Gaussian case the third cumulant is zero, in which case the conditional mean is $\tilde{\kappa}^i = \kappa^i + \kappa^{i,r}\bar{\kappa}_{r,s}(x^s - \kappa^s)$. This takes the form of a correction to the unconditional mean due to correlations between the two sets of variables. Non-Gaussianity imparts a quadratic correction to this proportional to the third cumulant. Likewise, the conditional covariance in the Gaussian case is $\tilde{\kappa}^{i,j} = \kappa^{i,j} - \kappa^{i,r} \bar{\kappa}_{r,s} \kappa^{s,j}$, which does not depend on the value of the variable on which we condition. This subtracts from the unconditional covariance a term accounting for correlations with the (fixed) variable $X^{(2)}$, i.e. the scatter in $X^{(1)}$ is not as great as it could be since the part correlated with $X^{(2)}$ must be held fixed. Non-Gaussianity provides a linear correction to the covariance, again proportional to the third cumulant.

\subsection{The conditional correlation function of unsmoothed fields}
\label{subsec:cond_corr}

We can now apply Equation~\eqref{eq:cond_cums} to the real-space dark-matter density field. We will neglect for now the effects of evolution and assume all the fields lie at the same redshift -- this is actually a reasonable approximation more generally since corrections from conditioning will only be significant for nearby structure. We wish to compute the first few cumulants of the density field $\delta(\mathbf{r}_i) \equiv \delta_i$ conditioned on the observer's local density field $\delta_0$, assumed to be at the origin of the coordinate system. To ensure that our truncated Edgeworth expansion is a good description of the true distribution will mean smoothing these density fields on some sufficiently large scale.

In the notation of Section~\ref{subsec:edge} we have $X^{i} = \delta_i$ and $X^r = \delta_0$, with $p-q = 1$. Since we deal with density contrasts we have $\kappa^i = \kappa^r = 0$, and by homogeneity (and neglect of evolution) we have $\langle \delta_i^2 \rangle = \langle \delta_0^2 \rangle \equiv \sigma^2$, where the dependence on the smoothing scale is left implicit for now. Plugging this into Equation~\eqref{eq:cond_cums} we get the conditional mean
\begin{equation}
  \langle \delta_i | \delta_0 \rangle = \langle \delta_i \delta_0 \rangle \frac{\delta_0}{\sigma^2} + \frac{1}{2}\left[ \langle \delta_i \delta_0^2 \rangle - \frac{\langle \delta_i \delta_0 \rangle}{\sigma^2} \langle \delta_0^3 \rangle \right] \left(\frac{\delta_0^2}{\sigma^4} - \frac{1}{\sigma^2}\right).
  \label{eq:cond_mean}
\end{equation}
The first term in Equation~\eqref{eq:cond_mean} is the Gaussian term expected from the discussion in Section~\ref{subsec:order}. The second term, proportional to $(\delta_0^2 - \sigma^2)$, is the leading-order correction from non-Gaussianity. Note that subsequent averaging of Equation~\eqref{eq:cond_mean} over $\delta_0$ yields zero, as required. The quantity $\langle \delta_i \delta_0 \rangle$ is just the correlation function $\xi(r_i)$.

Similarly, the conditional covariance is given by
\begin{align}
  \mathrm{cov}(\delta_i, \delta_j | \delta_0) &= \langle \delta_i \delta_j \rangle - \frac{\langle \delta_i \delta_0 \rangle \langle \delta_j \delta_0 \rangle}{\sigma^2} + \nonumber \\
  & \left[\langle \delta_i \delta_j \delta_0 \rangle - \frac{\langle \delta_i \delta_0 \rangle }{\sigma^2}\langle \delta_j \delta_0^2 \rangle - \right. \nonumber \\
  & \left. \frac{\langle \delta_j \delta_0 \rangle}{\sigma^2}\langle \delta_i \delta_0^2 \rangle + \frac{\langle \delta_i \delta_0 \rangle \langle \delta_j \delta_0 \rangle}{\sigma^4}\langle \delta_0^3 \rangle \vphantom{\frac12}\right]\frac{\delta_0}{\sigma^2}.
  \label{eq:cond_cov}
\end{align}
The first line of Equation~\eqref{eq:cond_cov} is the Gaussian expression, expected from the discussion in Section~\ref{subsec:order}. The other lines of Equation~\eqref{eq:cond_cov} are the leading-order corrections from non-Gaussianity and yield zero after subsequent averaging over $\delta_0$. They are proportional to various three-point functions of the local and remote density fields. In particular, it should be noted that the only term depending on the angular separation of $\mathbf{r}_i$ and $\mathbf{r}_j$ is $\langle \delta_i \delta_j \delta_0 \rangle \delta_0/\sigma^2$. We thus expect corrections at $l>0$ to come solely from this term. Note also the similarities and differences between the conditional correlation function of the density contrast and that of discrete tracers given in Ref.~\cite{Peebles}, where the conditional probability of finding two objects given the presence of a third follows almost immediately from the definition of the three-point function.

At leading order (tree-level) in standard perturbation theory we may write the three-point function of the unsmoothed density field (the expression for smoothed fields is cumbersome to write down -- we will address smoothed fields in the next section) in Einstein-de Sitter as~\cite{2002PhR...367....1B}
\begin{align}
  \langle \delta(\mathbf{r}_1)\delta(\mathbf{r}_2) \delta(\mathbf{r}_3) \rangle =& \, \frac{10}{7}\xi(r_{13})\xi(r_{23}) \nonumber \\
  & + \nabla \xi(\mathbf{r}_{13}) \cdot \nabla^{-1}\xi (\mathbf{r}_{23}) \nonumber \\
  & + \nabla \xi(\mathbf{r}_{23}) \cdot \nabla^{-1}\xi (\mathbf{r}_{13}) \nonumber \\
  & + \frac{4}{7}\left[ \nabla_a \nabla^{-1}_b \xi(\mathbf{r}_{13}) \right)\left( \nabla_a \nabla^{-1}_b \xi(\mathbf{r}_{23}) \right] \nonumber \\
  & + \mathrm{cyc.},
  \label{eq:xi3}
\end{align}
where $\mathbf{r}_{13} \equiv \mathbf{r}_1 - \mathbf{r}_3$ etc. and the terms involving the correlation function may be written in terms of the linear power spectrum $P(k)$ as\footnote{Our Fourier convention is such that $\delta(\mathbf{r}) = \int \frac{\mathrm{d}^3\mathbf{k}}{(2\pi)^3} \delta(\mathbf{k}) e^{i \mathbf{k} \cdot \mathbf{r}}$.}
\begin{align}
  \nabla \xi(r) &= i\int \frac{\mathrm{d}^3 \mathbf{k}}{(2\pi)^3} \mathbf{k} \, P(k) e^{i\mathbf{k} \cdot \mathbf{r}}, \label{eq:dxi}\\
  \nabla^{-1} \xi(\mathbf{r}) &= -i\int \frac{\mathrm{d}^3 \mathbf{k}}{(2\pi)^3} \frac{\mathbf{k}}{k^2} \, P(k) e^{i\mathbf{k} \cdot \mathbf{r}}, \label{eq:idxi}\\
  \nabla_a \nabla_b^{-1} \xi(\mathbf{r}) &= \int \frac{\mathrm{d}^3 \mathbf{k}}{(2\pi)^3} \frac{k_a k_b}{k^2} \, P(k) e^{i\mathbf{k} \cdot \mathbf{r}}. \label{eq:didxi}
\end{align}
To obtain $\langle \delta_i \delta_j \delta_0 \rangle$ we simply set $\mathbf{r}_3 = 0$ in Equation~\eqref{eq:xi3}. The integrals in Equations~\eqref{eq:dxi},~\eqref{eq:idxi} and~\eqref{eq:didxi} may be simplified by noting that by isotropy we must have
\begin{equation}
  \nabla_a \nabla_b^{-1}\xi(\mathbf{r}) = \xi(r)\frac{\delta_{ab}}{3} + \psi(r)\left(\hat{\mathbf{r}}_a \hat{\mathbf{r}}_b - \frac{\delta_{ab}}{3}\right),
\end{equation}
where $\psi(r) \equiv \frac{3}{2}(\hat{\mathbf{r}}_a \hat{\mathbf{r}}_b - \delta_{ab}/3)\nabla_a \nabla_b^{-1}\xi(\mathbf{r})$. This yields
\begin{equation}
  \psi(r) = -\int \frac{k^2\mathrm{d}k}{2\pi^2} P(k) j_2(kr).
  \label{eq:psi}
\end{equation}

Likewise, $\nabla \xi$ and $\nabla^{-1}\xi$ must both be proportional to $\hat{\mathbf{r}}$, so
\begin{align}
  \nabla \xi (\mathbf{r}) &= -\hat{\mathbf{r}}\int \frac{k^2\mathrm{d}k}{2\pi^2} kP(k) j_1(kr) \nonumber \\
  &= \hat{\mathbf{r}}\int \frac{k^2\mathrm{d}k}{2\pi^2} kP(k) j_0'(kr) \nonumber \\
  &=  \hat{\mathbf{r}} \xi'(r),
  \label{eq:dxi_simple}
\end{align}
and
\begin{align}
  \nabla^{-1} \xi (\mathbf{r}) &= \hat{\mathbf{r}}\int \frac{k^2\mathrm{d}k}{2\pi^2} \frac{P(k)}{k} j_1(kr) \nonumber \\
  &\equiv \hat{\mathbf{r}}\Omega(r).
  \label{eq:omega}
\end{align}

Plugging the above expressions into Equation~\eqref{eq:cond_cov} gives the conditional correlation function at tree-level in Einstein de-Sitter as
\begin{align}
  &\mathrm{cov}(\delta_i, \delta_j | \delta_0) = \xi(d) - \frac{\xi(r_i)\xi(r_j)}{\sigma^2} + \nonumber \\
  &\left\{ \frac{34}{21}\xi(d)\left[\xi(r_i) + \xi(r_j)\right] \right. \nonumber \\
  &\left. + \left[\xi'(r_i)\Omega(r_j) + \xi'(r_j)\Omega(r_i)\right]\cos \beta \right. \nonumber \\
  &\left. - \left[\xi'(d)\Omega(r_i) + \xi'(r_i)\Omega(d)\right]\cos \phi \nonumber \right. \\
  & \left. + \left[\xi'(d)\Omega(r_j) + \xi'(r_j)\Omega(d)\right]\cos \alpha \right. \nonumber \\
  &\left. + \frac{4}{7}\left[\psi(r_i)\psi(r_j)\left(\cos^2\beta - \frac{1}{3}\right) \right. \right. \nonumber \\
    & \left. \left. + \psi(d)\psi(r_i)\left(\cos^2\phi - \frac{1}{3}\right) \right. \right. \nonumber \\
    & \left. \left. + \psi(d)\psi(r_j)\left(\cos^2\alpha - \frac{1}{3}\right) \right] \right. \nonumber \\
  &\left. - \frac{\xi(r_j)}{\sigma^2}\left[\frac{34}{21}\xi(r_i)^2 + \xi'(r_i)\Omega(r_i) + \frac{8}{21}\psi(r_i)^2\right] \right. \nonumber \\
  & \left. - \frac{\xi(r_i)}{\sigma^2}\left[\frac{34}{21}\xi(r_j)^2 + \xi'(r_j)\Omega(r_j) + \frac{8}{21}\psi(r_j)^2\right] \right\}\frac{\delta_0}{\sigma^2}.
  \label{eq:cond_cov_tree}
\end{align}
where we defined the angles $\cos \beta = \hat{\mathbf{r}}_i \cdot \hat{\mathbf{r}}_j$, $\cos\phi = \hat{\mathbf{r}}_i \cdot \widehat{\mathbf{r}_j - \mathbf{r}_i}$, $\cos\alpha = \hat{\mathbf{r}}_j \cdot \widehat{\mathbf{r}_j - \mathbf{r}_i}$. Evolution could be accounted for by replacing the correlation functions with their appropriate unequal-time counterparts, i.e. with the appropriate linear growth factors. The term in braces is the non-Gaussian correction, linear in the local density fluctuation.

The correction terms in Equation~\eqref{eq:cond_cov_tree} have been grouped according to their angular dependence. Recall that in standard Eulerian perturbation theory, the second-order density field can be written as the sum of a monopolar density-squared term, a dipolar shift term, and a quadrupolar tidal term. The second line of Equation~\eqref{eq:cond_cov_tree} is the product of the monopole second-order density at $\mathbf{r}_i$ with the linear densities at $\mathbf{r}_j$ and the origin, with a corresponding term for $\mathbf{r}_i$ by symmetry (the corresponding term with the second-order density at the origin has been cancelled by the remaining terms in square brackets in Equation~\eqref{eq:cond_cov}). The third, fourth, and fifth lines are the products of the dipole second-order densities at the origin, $\mathbf{r}_i$, and $\mathbf{r}_j$  respectively with the linear densities at the other points, and the sixth, seventh, and eighth lines are the equivalent terms for the tidal part of the second-order density. Finally, the ninth and tenth lines are the remaining terms in square brackets in Equation~\eqref{eq:cond_cov} which have not cancelled.

Equation~\eqref{eq:cond_cov_tree} is not particularly useful since none of the fields involved has been smoothed. In particular, no smoothing scale has been specified for $\delta_0$. The real-space three-point function for smoothed fields is rather complicated and better described in Fourier space, so we defer discussion of smoothing to the next section where we compute the conditional angular power spectrum. Nevertheless, we can gain some insight from this expression. Consider the case where only long-wavelength modes contribute to the local density field, such that it can be taken as linear (we will tighten up this statement in the next section where we consider smoothed fields). Then we can neglect the third, sixth, ninth and tenth lines of Equation~\eqref{eq:cond_cov_tree}. If only wavenumbers having $k < k_{\mathrm{max}}$ contribute to $\delta_0$ then when $r \ll k^{-1}_{\mathrm{max}}$ we have $\xi(r) \approx \sigma^2$ and $\Omega(r) \approx r \sigma^2/3$. We also have $\xi'(r) \Omega(d) \ll \xi'(d)\Omega(r)$ because the integrand of $\xi'(r)$ is suppressed by the restriction $k < k_{\mathrm{max}}$ while the integrand of $\Omega(r)$ is enhanced -- see Equations~\eqref{eq:dxi_simple} and~\eqref{eq:omega}. The tidal terms proportional to $\psi(r)$ are quadratic in $r$ when $r \ll k^{-1}_{\mathrm{max}}$, so the seventh and eighth lines of Equation~\eqref{eq:cond_cov_tree} are suppressed by factors $\sim (r k_{\mathrm{max}})^2 $ and can hence be neglected. This leaves us with
\begin{align}
  \mathrm{cov}(\delta_i, \delta_j | \delta_0) \approx & \, \xi(d) - \sigma^2 \nonumber \\
  & + \left \{ \frac{68}{21}\xi(d) + [r_j \cos \alpha - r_i \cos \phi]\frac{\xi'(d)}{3} \right \} \delta_0 \nonumber \\
  = & \, \xi(d) - \sigma^2 + \left[ \frac{68}{21}\xi(d) + d\frac{\xi'(d)}{3} \right ] \delta_0 \nonumber \\
  &(k_{\mathrm{max}} \ll k_{\mathrm{NL}}, r_i^{-1}, r_j^{-1}),
  \label{eq:approx_cond_cov}
\end{align}
where the third line follows from the second by the definitions of the angles $\alpha$ and $\phi$. Equation~\eqref{eq:approx_cond_cov} is the Gaussian conditional correlation plus a term precisely matching the approximate conditional correlation function derived in Section~\ref{subsec:order} using effective curvature arguments. This should come as no surprise of course but provides a useful check on Equation~\eqref{eq:cond_cov_tree}.
  
\subsection{The conditional angular power spectrum of smoothed fields}
\label{subsec:cond_cls}

In the previous section we applied the general conditional covariance expression Equation~\eqref{eq:cond_cov} to the real-space dark matter field, which allowed us to quickly derive the conditional correlation of unsmoothed fields Equation~\eqref{eq:cond_cov_tree}. This expression is not particularly useful however, since no smoothing scale has been specified for $\delta_0$, the local density contrast. Ultimately we expect the corrections from conditioning will be most significant for gravitational lensing observables, which inevitably pick up contributions from nearby structure in projection. This suggests we need not worry too much about smoothing the two remote density fields, since for lensing we will ultimately be effectively replacing them with the gravitational potential.

Smoothing is most easily implemented in Fourier space, and since the non-Gaussian terms which are the focus of this work are the only contribution for $l>0$ we will now compute the conditional angular power spectrum of density fluctuations. Theoretical modelling of cosmological statistics usually starts with a Fourier-space expression, which provides further motivation for working in spherical-harmonic space. The derivation is slightly involved, and the reader only interested in the final expression may skip to Equation~\eqref{eq:bll0_s}.

Since conditioning on $\delta_0$ introduces no preferred direction, the conditional density field must be statistically isotropic\footnote{This can be seen at tree-level from Equation~\eqref{eq:cond_cov_tree}, where the conditional correlation function depends only on $r_i$, $r_j$, and $\cos \beta$.}. With $\delta_{lm}(r)$ the spherical multipoles of the density field at $\mathbf{r}$, we define the conditional angular power spectrum $\tilde{C}_l(r_i, r_j)$ as
\begin{equation}
  \mathrm{cov}(\delta_{lm}(r_i), \delta_{l'm'}(r_j) | \delta_0 ) = (-1)^{m'} \delta^K_{ll'} \delta^K_{m-m'} \tilde{C}_l(r_i,r_j),
\end{equation}
where $\delta^K_{ab}$ is the Kronecker delta and where $\tilde{C}_l(r_i, r_j)$ is symmetric in its arguments and can be written in terms of the real-space conditional correlation function as
\begin{align}
  &\tilde{C}_l(r_i,r_j) = \nonumber \\
  & \, \, \int \mathrm{d}^2 \hat{\mathbf{r}}_i \, Y_{lm}^*(\hat{\mathbf{r}}_i) \int \mathrm{d}^2 \hat{\mathbf{r}}_j \, Y_{lm}(\hat{\mathbf{r}}_j) \mathrm{cov}( \delta(\mathbf{r}_i), \delta(\mathbf{r}_j) | \delta_0 ).
\end{align}
Substituting in Equation~\eqref{eq:cond_cov} and noting that $\delta_{00}(0) = \sqrt{4\pi}\delta_0$ we find that
\begin{align}
  &\tilde{C}_l(r_i,r_j) = C_l(r_i,r_j) - \frac{4\pi \xi(r_i)\xi(r_j)}{\sigma^2}\delta^K_{l0} \nonumber \\
  & + \left[B^{m -m 0}_{l l 0}(r_i,r_j,0)(-1)^m - \xi(r_i)B^{000}_{000}(r_j,0,0)\delta^K_{l0} \right. \nonumber \\
    & \left. - \xi(r_j)B^{000}_{000}(r_i,0,0)\delta^K_{l0} + \xi(r_i)\xi(r_j)B^{000}_{000}(0,0,0)\delta^K_{l0}\right] \nonumber \\
  & \, \, \times \frac{\delta_0}{\sigma^2\sqrt{4\pi}},
  \label{eq:cond_cl_general}
\end{align}
where we have introduced the bispectrum of the multipoles defined by $\langle \delta_{l_1m_1}(r_1)\delta_{l_2m_2}(r_1) \delta_{l_3m_3}(r_3) \rangle \equiv B^{m_1m_2m_3}_{l_1l_2l_3}(r_1, r_2, r_3)$. Note that $C_0(r_i,0) = 4\pi\xi(r_i)$.

The $m$-dependence of the bispectrum follows from isotropy, and leads to the definition of the reduced bispectrum $b_{l_1l_2l_3}(r_1,r_2,r_3)$ as
\begin{align}
  &B^{m_1m_2m_3}_{l_1l_2l_3}(r_1 r_2 r_3) = \sqrt{\frac{(2l_1 + 1)(2l_2 + 1)(2l_3 + 1)}{4\pi}} \nonumber \\
  & \, \, \times \ThreeJSymbol{l_1}{0}{l_2}{0}{l_3}{0} \ThreeJSymbol{l_1}{m_1}{l_2}{m_2}{l_3}{m_3} b_{l_1l_2 l_3}(r_1,r_2,r_3),
\end{align}
where the terms in parentheses are the Wigner 3$j$ symbols. From this it follows that the conditional angular power spectrum is
\begin{align}
  &\tilde{C}_l(r_i,r_j) = C_l(r_i,r_j) - \frac{4\pi \xi(r_i)\xi(r_j)}{\sigma^2}\delta^K_{l0} \nonumber \\
  & + \left[b_{l l 0}(r_i,r_j,0) - \xi(r_i)b_{000}(r_j,0,0)\delta^K_{l0} \right. \nonumber \\
  & \left. - \xi(r_j)b_{000}(r_i,0,0)\delta^K_{l0} + \xi(r_i)\xi(r_j)b_{000}(0,0,0)\delta^K_{l0}\right]\frac{\delta_0}{4\pi\sigma^2},
  \label{eq:cond_cl_reduced}
\end{align}
Therefore, for $l>0$ the leading-order correction to the angular power spectrum is $b_{ll0}(r_i,r_j,0)\delta_0/(4\pi\sigma^2)$.

It is worth noting at this point that we have assumed that the density field is accessible over the full sky in order to obtain the full set of spherical multipoles. In reality we observe the density field in a finite survey window, which complicates the calculation of the conditional power spectrum since the window induces mixing between Fourier modes. One particular feature of the finite-sky case relevant to the discussion is the use of a local average density instead of a global average in defining the density contrast. As discussed in Ref.~\cite{2012JCAP...04..019D}, the \emph{measured} density contrast in the survey window is $\delta_M(\hat{\mathbf{n}}) = [\delta(\hat{\mathbf{n}}) - \delta_L]/(1 + \delta_L)$, where $\delta_L$ is the density contrast averaged over the survey patch mostly having contributions from super-survey modes. Since the local density field correlates with $\delta(\hat{\mathbf{n}})$ in exactly the same way as $\delta_L$, to leading order this measured field is uncorrelated with $\delta_0$ and hence in the Gaussian case there is no correction from conditioning, even for $l=0$. The same argument in fact holds on the full sky, but there the ambiguity between using the ensemble-averaged mean density or the realization-dependent measured mean density to define density contrasts only impacts the $l=0$ mode. On the cut sky one needs to be careful about which mean one is using, but since in practice we are interested in lensing we will consistently define density contrasts with respect to the FRW ensemble-averaged density.

To keep things as symmetric for as long as possible we will first compute the general reduced bispectrum $b_{l_1 l_2 l_3}(r_1, r_2, r_3)$. This was first done in Ref.~\cite{1999PhRvD..59j3001S}, although there are some missing factors of $i$ which are corrected in Ref.~\cite{2016JCAP...01..016D}. We repeat the derivation here in a slightly clearer fashion and incorporate smoothing.

Writing $\delta(\mathbf{r}) = \delta^{(1)}(\mathbf{r}) + \delta^{(2)}(\mathbf{r}) + ...$, the bispectrum of smoothed fields follows from writing the density field multipoles in terms of the Fourier modes $\delta(\mathbf{k})$ and reads
\begin{align}
  &B^{m_1 m_2 m_3}_{l_l l_2 l_3}(r_1, r_2, r_3) = \nonumber \\
  & \, \, \int \frac{k_1^2 \mathrm{d}k_1}{2\pi^2} W(k_1R)\int \frac{k_2^2 \mathrm{d}k_2}{2\pi^2} W(k_2R)\int \frac{k_3^2 \mathrm{d}k_3}{2\pi^2} W(k_3R) \nonumber \\
  & \, \, \int \mathrm{d}^2\hat{\mathbf{k}}_1\int \mathrm{d}^2\hat{\mathbf{k}}_2\int \mathrm{d}^2\hat{\mathbf{k}}_3 \, i^{l_1+l_2+l_3}\nonumber \\
  & \, \, \times j_{l_1}(k_1r_1)j_{l_2}(k_2r_2)j_{l_3}(k_3r_3) Y_{l_1m_1}^*(\hat{\mathbf{k}}_1)Y_{l_2m_2}^*(\hat{\mathbf{k}}_2)Y_{l_3m_3}^*(\hat{\mathbf{k}}_3) \nonumber \\
  & \, \, \times  \langle \delta^{(2)}(\mathbf{k}_1)\delta^{(1)}(\mathbf{k}_2)\delta^{(1)}(\mathbf{k}_3)\rangle + \mathrm{cyc.},
  \label{eq:Blm}
\end{align}
where $W(kR)$ is the smoothing kernel, which we assume is isotropic.

Now, let's focus on the term in Equation~\eqref{eq:Blm} for which the non-linear field is $\delta(\mathbf{r}_1)$. The tree-level Fourier space bispectrum in Einstein de-Sitter (although the cosmology-dependence is very weak) is~\cite{2002PhR...367....1B}
\begin{align}
  &\langle \delta^{(2)}(\mathbf{k}_1)\delta^{(1)}(\mathbf{k}_2)\delta^{(1)}(\mathbf{k}_3) \rangle  = \nonumber \\
  & \, \, 2(2\pi)^3 F_2(\mathbf{k}_2,\mathbf{k}_3)P(k_2)P(k_3)\delta^D(\mathbf{k}_1 + \mathbf{k}_2 + \mathbf{k}_3).
\end{align}
We now write the $F_2$ kernel as
\begin{align}
  F_2(\mathbf{k}_2,\mathbf{k}_3) &= \sum_Lf_L(k_2, k_3) \mathcal{L}_L(\hat{\mathbf{k}}_2 \cdot \hat{\mathbf{k}}_3) \nonumber \\
  &= 4\pi \sum_L\frac{f_L(k_2, k_3)}{2L+1}\sum_MY_{LM}(\hat{\mathbf{k}}_2)Y^*_{LM}(\hat{\mathbf{k}}_3)
\end{align}
where $\mathcal{L}_L$ is a Legendre polynomial. This implies that $f_0(k_2,k_3) = 17/21$, $f_1(k_2,k_3) = (k_2/k_3 + k_3/k_2)/2$, $f_2(k_2,k_3) = 4/21$, and $f_L(k_2,k_3) = 0$ for $L>2$.
%
%
%

Now we write the Dirac delta function as the Fourier transform of unity and expand its Fourier exponential in spherical harmonics. The integral over $\hat{\mathbf{k}}_1$ in Equation~\eqref{eq:Blm} for our term is trivial and enforces $l = l_1$ and $m = -m_1$. The integrals over $\hat{\mathbf{k}}_2$ and $\hat{\mathbf{k}}_3$ involve three spherical harmonics and so can be written as products of $3j$ symbols. The same is true of the integral over $\hat{\mathbf{r}}$. This leaves a sum over the product of six $3j$ symbols, and the remaining integrals over $k_1$, $k_2$, $k_3$, and $r$. The three $3j$ symbols with non-zero $m$-arguments can be summed over $m$ using Equation 8.7.3.12 of Ref.~\cite{Varsh} and written as the product of a $6j$ symbol and a $3j$ symbol -- this $3j$ symbol contains all the $m$-dependence of the bispectrum and is consistent with isotropy. Following the notation of Ref.~\cite{1999PhRvD..59j3001S} and Ref.~\cite{2016JCAP...01..016D} we define the geometric quantities
\begin{equation}
  g_{l_1 l_2 l_3} \equiv \sqrt{\frac{(2l_1 + 1)(2l_2 + 1)(2l_3 + 1)}{4\pi}} \ThreeJSymbol{l_1}{0}{l_2}{0}{l_3}{0}
\end{equation}
and
\begin{align}
  &I^{l_1 l_2 l_3}_{l l' l''} \equiv \sqrt{(4\pi)^3(2l_1 + 1)(2l_2 + 1)(2l_3 + 1)} \nonumber \\
  & \, \, \times \ThreeJSymbol{l}{0}{l''}{0}{l_1}{0}\ThreeJSymbol{l'}{0}{l}{0}{l_2}{0}\ThreeJSymbol{l''}{0}{l'}{0}{l_3}{0}
\end{align}
and finally
\begin{equation}
  Q^{l_1l_2l_3}_{ll'l''} \equiv I^{l_1 l_2 l_3}_{l l' l''} \SixJSymbol{l_1}{l'}{l_2}{l''}{l_3}{l}(-1)^{l+l'+l''},
\end{equation}
where the term in braces is a Wigner 6$j$ symbol.

Making all these replacements, using the explicit expressions for the $f_L$ coefficients and then using the definition of the reduced bispectrum gives, finally
\begin{align}
  &b_{l_1 l_2 l_3}(r_1, r_2, r_3) = 4\pi \frac{\{l_1 \, l_2 \, l_3\}}{g_{l_1 l_2 l_3}} \nonumber \\
  &\, \, \times \int \frac{k_1^2 \mathrm{d}k_1}{2\pi^2} W(k_1R)\int \frac{k_2^2 \mathrm{d}k_2}{2\pi^2} W(k_2R)\int \frac{k_3^2 \mathrm{d}k_3}{2\pi^2} W(k_3R) \nonumber \\
  & \, \, \times (-i)^{l_2 + l_3} j_{l_1}(k_1r_1)j_{l_2}(k_2r_2)j_{l_3}(k_3r_3)\int r^2 \mathrm{d}r \, j_{l_1}(k_1r) \nonumber \\
  &\, \, \times \sum_{l',l''} i^{l'+l''}(2l'+1)(2l''+1)j_{l'}(k_2r)j_{l''}(k_3r) 2P(k_2)P(k_3) \nonumber \\
  &\, \, \times  \left[\frac{17}{21}Q^{l_1 l_2 l_3}_{l' 0 l''} + \frac{1}{2}\left(\frac{k_2}{k_3} + \frac{k_3}{k_2}\right)Q^{l_1 l_2 l_3}_{l' 1 l''} + \frac{4}{21}Q^{l_1 l_2 l_3}_{l' 2 l''}\right] \nonumber \\
  &\, \, + \mathrm{sym.},
  \label{eq:reduced_b}
\end{align}
where $\{l_1 \, l_2 \, l_3\} = 1$ if the symmetries of $g_{l_1 l_2 l_3}$ are enforced (i.e. the triangle conditions and $l_1 + l_2 + l_3$ equalling an even integer), and zero otherwise. Note that Equation~\eqref{eq:reduced_b} agrees with the expression in Ref.~\cite{2016JCAP...01..016D}, which can be shown by using the symmetries of the $3j$ and $6j$ symbols and some relabelling. Note that the power spectra are really unequal-time power spectra, and we should replace $P(k_2) \rightarrow P(k_2; z_2, z_1)$ and $P(k_3) \rightarrow P(k_3; z_3, z_1)$ in Equation~\eqref{eq:reduced_b}.

Now, two of the integrals in Equation~\eqref{eq:reduced_b} can be done analytically using the orthogonality of spherical Bessel functions if we assume $W(k_1R) = 1$, i.e. \emph{if we leave the non-linear field unsmoothed in each term}. Since we will want to at least smooth the local density field, this suggests that we will have to neglect any non-linearity in $\delta_0$. With $R$ the smoothing scale of the local density, this implies that we must choose $R$ sufficiently large to suppress non-linearity in $\delta_0$. The resulting smoothed field is $\delta_0(R)$. Since we wish to maintain non-linearity in the non-local fields, we can only make further analytic progress if we leave these fields unsmoothed. This is acceptable if we wish to compute lensing spectra, since the relevant density fields are unsmoothed in this case.
%
%
%

We now take the limit of Equation~\eqref{eq:reduced_b} when $\delta(\mathbf{r}_3) = \delta_0$, the local density field smoothed on scale $R$. We set $r_1 = r_i$, $r_2 = r_j$, $r_3 = 0$, $l_1 = l$, $l_2 = l$, and $l_3 = 0$. We'll continue to work with the term having $\delta(\mathbf{r}_1)$ as a non-linear field, with the other two fields linear. The triangle condition is clearly met, and $\{l \, l\, 0\} = 1$. We also have $g_{ll0} = (-1)^l\sqrt{(2l+1)/4\pi}$. The coefficients $Q^{ll0}_{l'Ll''}$ can be computed straightforwardly for $L=0,1,2$ and are
\begin{align}
  Q^{ll0}_{l'0l''} &= \delta^K_{l''0} (4\pi)^{3/2}\frac{(-1)^l}{\sqrt{2l+1}}\delta^K_{l'l},  \nonumber\\
  Q^{ll0}_{l'1l''} &= \delta^K_{l''1}\frac{(4\pi)^{3/2}(-1)^l}{3\sqrt{2l+1}} \left( \frac{l+1}{2l+3}\delta^K_{l',l+1} + \frac{l}{2l-1}\delta^K_{l',l-1} \right), \nonumber\\
  Q^{ll0}_{l'2l''} &= \delta^K_{l''2}\frac{(4\pi)^{3/2}(-1)^l}{5\sqrt{2l+1}}\left[\frac{3(l+1)(l+2)}{2(2l+3)(2l+5)}\delta^K_{l',l+2} \right. \nonumber \\
  & \, \, \left. + \frac{l(l+1)}{(2l-1)(2l+3)}\delta^K_{l',l} + \frac{3(l-1)l}{2(2l-3)(2l-1)}\delta^K_{l',l-2} \right].
\end{align}
%


Substituting these coefficients into Equation~\eqref{eq:reduced_b}, using the recursion relations for the spherical Bessel functions and their derivatives, and neglecting non-linearity in $\delta_0(R)$, we finally obtain the conditional angular power spectrum for $l>0$ as $\tilde{C}_l(r_i,r_j) \approx C_l(r_i,r_j) + \Delta C_l(r_i,r_j)$ with 
\begin{widetext}
\begin{align}
  \Delta C_l(r_i,r_j)&= \, 4\pi\int \frac{k^2\mathrm{d}k}{2\pi^2}P(k;z_i,z_j) j_l(kr_j) \nonumber \\
  &\times \left\{ \frac{34}{21}\xi_R(r_i)j_l(kr_i) + \left[k\Omega_R(r_i) - \frac{\xi_R'(r_i)}{k}\right]j_l'(kr_i) - \frac{8}{21}\psi_R(r_i)\left[\frac{3}{2}j_l''(kr_i) + \frac{1}{2}j_l(kr_i)\right]\right\}\frac{\delta_0(R)}{\sigma^2(R)}  + (r_i \leftrightarrow r_j).
  \label{eq:bll0_s}
\end{align}
\end{widetext}

\begin{figure*}
  \includegraphics[width=\columnwidth]{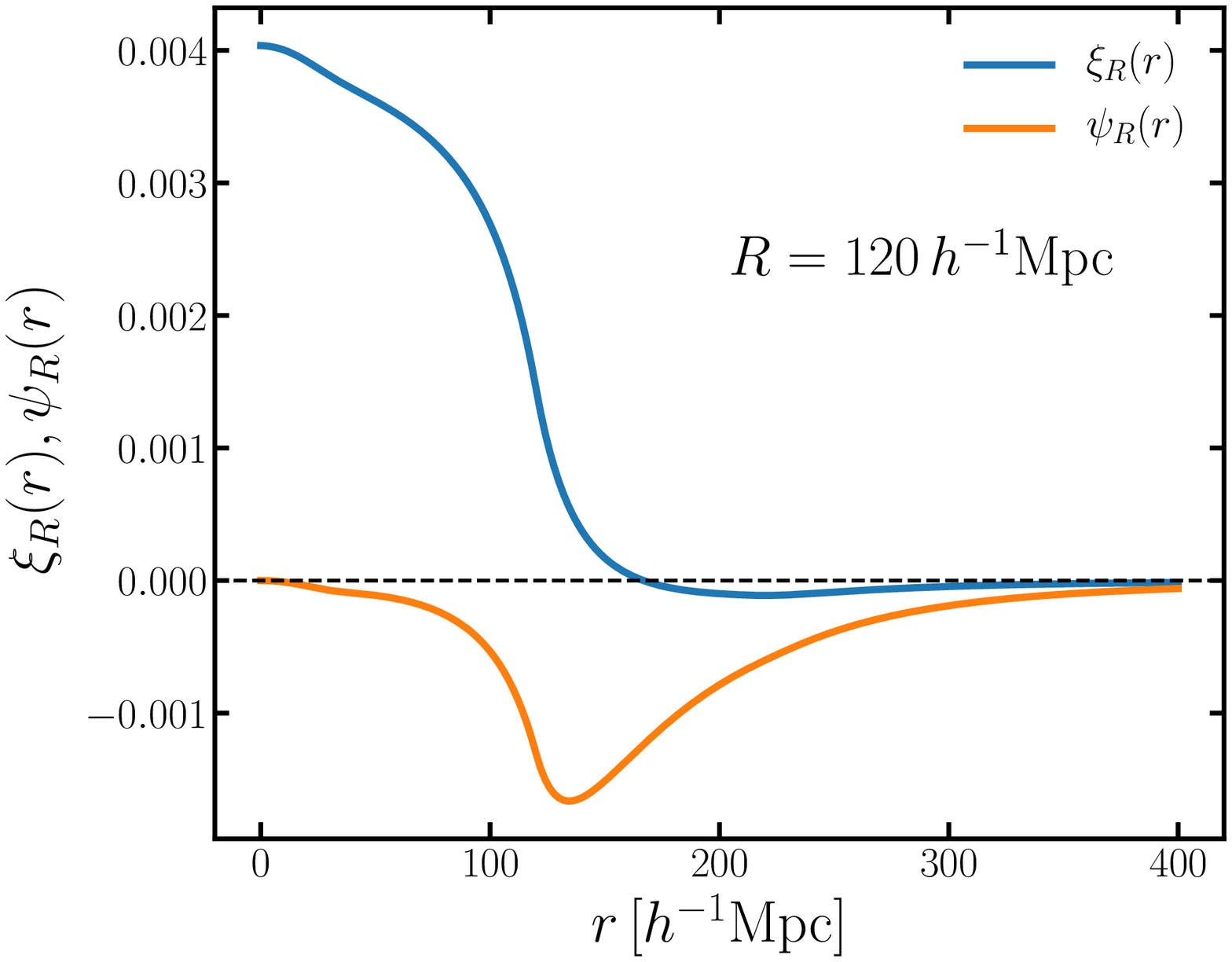}
  \includegraphics[width=\columnwidth]{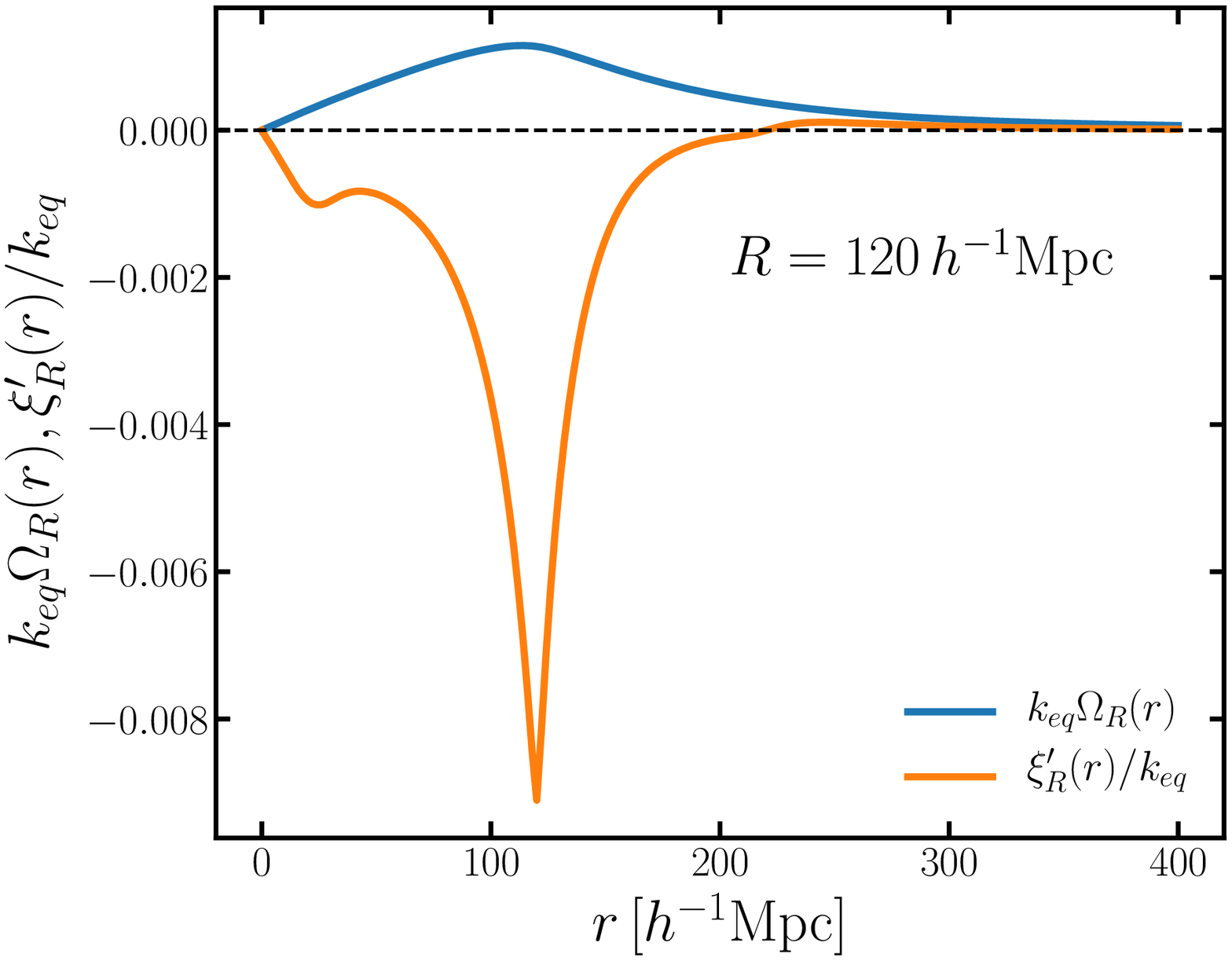}
  \caption{\emph{Left panel}: The functions $\xi_R(r)$ (upper blue) and $\psi_R(r)$ (lower orange) defined in the text for a smoothing scale of $R = 120 \, h^{-1} \, \mathrm{Mpc}$. \emph{Right panel}: The functions $\Omega_R(r)$ and $\xi_R'(r)$ defined in the text, normalised to the equality scale $k_{eq}$ and for $R = 120 \, h^{-1} \, \mathrm{Mpc}$. The dashed horizontal line denotes zero correlation. Both panels assume a linear matter power spectrum.}
  \label{fig:NG_corrs}
\end{figure*}

The correlation functions of Section~\ref{subsec:cond_corr} reappear in Equation~\eqref{eq:bll0_s} in smoothed form as
\begin{align}
  \xi_R(r) &= \int \frac{k^2\mathrm{d}k}{2\pi^2}P(k;z,0) j_0(kr)W(kR), \label{eq:xiR}\\
  \xi_R'(r) &= -\int \frac{k^2\mathrm{d}k}{2\pi^2}P(k;z,0) k j_1(kr)W(kR),\label{eq:dxiR}\\
  \Omega_R(r) &= \int \frac{k^2\mathrm{d}k}{2\pi^2}\frac{P(k;z,0)}{k} j_1(kr)W(kR),\label{eq:omegaR}\\
  \psi_R(r) &= -\int \frac{k^2\mathrm{d}k}{2\pi^2}P(k;z,0) j_2(kr)W(kR).\label{eq:psiR}
\end{align}
Note that $\sigma^2(R) \neq \xi_R(0)$, since $\sigma^2(R)$ has $W(kR)^2$ in its integrand rather than $W(kR)$.

Equation~\eqref{eq:bll0_s} is the first main result of this work, representing the leading-order correction to the angular power spectrum of density fluctuations in spherical shells for $l>0$ due to conditioning on the local smoothed density contrast. One may verify that this expression is consistent with the real-space three-point function in Equation~\eqref{eq:cond_cov_tree}. To show this, note that $\partial d/\partial r_i = -\cos \phi$ and $\partial d/\partial r_j = \cos \alpha$, and that $j_1$ and $j_2$ can be written as derivatives of $j_0$. The agreement then follows from taking second derivatives of $d$ with respect to $r_i$ and $r_j$.

So far, the only source of stochasticity in deriving Equation~\eqref{eq:bll0_s} has been that of the density field itself, i.e. cosmic variance. In practice the fields also contain observational (including Poisson) noise. This can be easily incorporated by replacing all power spectra and variances in the above equations with their noisy versions. In particular, we can in practice only make a noisy estimate of the local density field. In this case, the variance $\sigma(R)$ should be replaced by $\sqrt{\sigma^2(R) + \sigma^2_N(R)}$, where $\sigma^2_N(R)$ is the smoothed noise variance. $\delta_0(R)$ is then a noisy estimate of the local density, for example that arising from a point estimate such as a maximum-likelihood estimator. To keep our results as general as possible, we will neglect noise and instead account for it by quoting uncertainties in $\delta_0(R)$ or $\nu(R) = \delta_0(R)/\sigma(R)$ which should bracket the typical corrections $\Delta C_l$.

In Figure~\ref{fig:NG_corrs} we plot the four correlation functions defined in Equation~\eqref{eq:xiR}--\eqref{eq:psiR} for our preferred smoothing scale $R = 120 \, h^{-1}\mathrm{Mpc}$. All the correlation functions except $\xi_R(r)$ go to zero as $r \rightarrow 0$, and all go to zero for $r \gg R$. The result is that $\xi_R'(r)$, $\Omega_R(r)$ and $\psi_R(r)$ have support mostly around $r \approx R$, whereas $\xi_R(r)$ is large for $r \lesssim R$ and strongly suppressed for $r \gtrsim R$.

\begin{figure*}
  \includegraphics[width=\columnwidth]{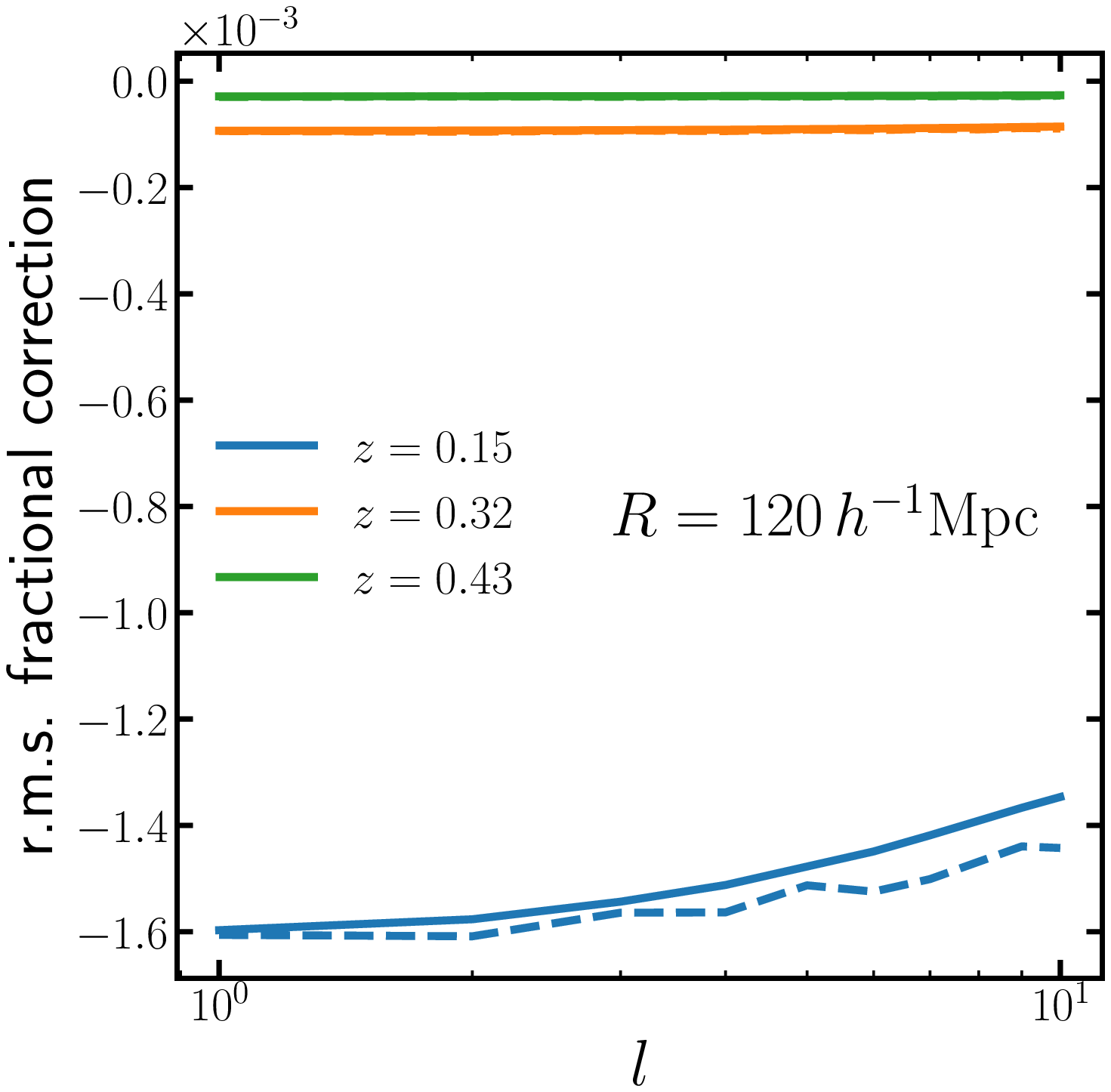}
  \includegraphics[width=\columnwidth]{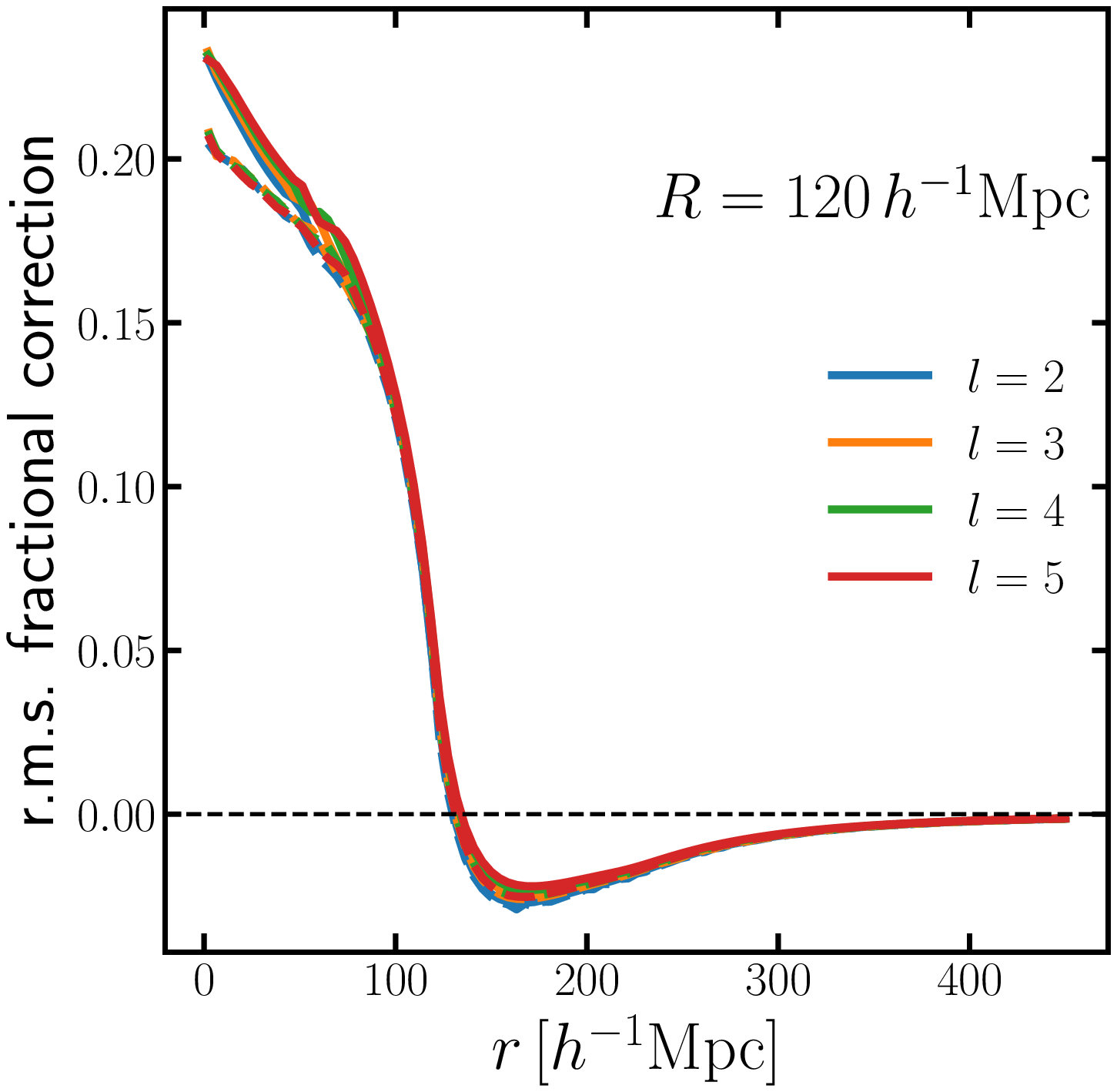}
  \caption{\emph{Left panel}: R.m.s. fractional correction to the $l>0$ angular power spectrum of matter density fluctuations on spherical shells at $z=0.15$ (blue lower curve), $z=0.32$ (orange middle curve) and $z=0.43$ (green upper curve), for $R = 120 \, h^{-1} \, \mathrm{Mpc}$. Dashed curves use a matter power spectrum corrected with \textsc{halofit}. The correction due to our own local density can be obtained by multiplying these curves by $\nu(R) = \delta_0(R)/\sigma(R)$, with our \citetalias{2015MNRAS.450..317C} estimate being  $\nu(120 \, h^{-1} \, \mathrm{Mpc}) \approx 0.85 \pm 0.53$. Corrections at these distances are at the sub-percent level and well within the (conditional) cosmic variance on all angular scales. \emph{Right panel}: Same as left panel but as a function of the radial distances of the spherical shells and for the largest few angular scales. All the curves lie roughly on top of each other. The radial behaviour seen here roughly matches that of the $\xi_R(r)$ function plotted in Figure~\ref{fig:xi_Rs}. Corrections can reach $\sim 10\%$ for nearby shells but are still within the conditional cosmic variance, which is roughly $\sqrt{2/(2l+1)}$ in units of the fractional difference.}
  \label{fig:diff_cond_Cls}
\end{figure*}

In Figure~\ref{fig:diff_cond_Cls} we plot the fractional correction to the angular auto power spectrum of density fluctuations in units of the normalized local fluctuation $\nu(R) = \delta_0(R)/\sigma(R)$, i.e. the r.m.s. fractional correction. We set the smoothing scale equal to $R = 120 \, h^{-1}\mathrm{Mpc}$, as discussed in Section~\ref{subsec:order}. The correction due to our own local density can be obtained by multiplying the curves in Figure~\ref{fig:diff_cond_Cls} by $\nu(R)$, where our estimate from the \citetalias{2015MNRAS.450..317C} density field is $\nu(120 \, h^{-1} \, \mathrm{Mpc}) \approx 0.85 \pm 0.53$ -- we choose to display results in units of $\nu(R)$ so that any estimate of the local density contrast can be inserted. In the left panel we plot results for three spherical shells located at $z=0.15$, $z=0.32$, and $z=0.43$, corresponding to the lower, effective, and upper redshifts of the BOSS LOWZ galaxy sample~\cite{2016MNRAS.455.1553R}, as a function of angular multipole. In the right panel we plot the radial dependence for a few choices of $l$.

The typical size of the correction to the $C_l$ are $\sim 10^{-3}$ for the BOSS LOWZ redshift shells, i.e. at the sub-percent level. At the effective redshift of LOWZ the corrections are smaller, at the $\sim 10^{-4}$ level and hence negligible for $\nu(R) \sim 1$ as suggested by the 2M++ catalogue. As shown in the left panel of Figure~\ref{fig:diff_cond_Cls}, the largest effects are at the largest angular scales, although the dependence on $l$ is quite weak. Note also that the sign of the correction is \emph{negative}, i.e. for positive $\delta_0(R)$ conditioning on the local density \emph{suppresses} the angular power spectrum for these redshifts.

The correction from conditioning is larger for closer redshift shells, as expected. The right panel of Figure~\ref{fig:diff_cond_Cls} shows that for $r \lesssim 100 \, h^{-1}\mathrm{Mpc}$ (i.e. $z \lesssim 0.03$) the corrections can be $\gtrsim 10\%$, roughly independent of $l$ (in agreement with the left panel). The radial dependence shown here roughly follows that of $\xi_R(r)$, i.e. the green curve in Figure~\ref{fig:xi_Rs}. For $r \gtrsim R$ the correction quickly becomes sub-percent.

Since our expression for the conditional power spectrum is only valid at second order in perturbation theory, we need to be careful to avoid too much sensitivity to very non-linear scales. To check this we compute Equation~\eqref{eq:bll0_s} using both the linear power spectrum (formally correct at this order in perturbation theory) and a non-linear power spectrum computed using the \textsc{halofit} correction of Ref.~\cite{2012ApJ...761..152T}, shown as the dashed curves in Figure~\ref{fig:diff_cond_Cls}. The impact of non-linear scales is non-negligible for all terms but is generally at the 10\% level. At $l \gtrsim 10$ the sensitivity to non-linear scales is more severe -- the small oscillations seen in the left panel of Figure~\ref{fig:diff_cond_Cls} are numerical artefacts arising from truncation of the $k$-integral in Equation~\eqref{eq:bll0_s} at $k_{\mathrm{max}} = 2.5 \, h^{-1}\mathrm{Mpc}$. The sensitivity of the non-Gaussian correction $\Delta C_l$ to small scales is greater than for the unconditional angular power spectrum, so we must take care to check that our results are not too sensitive to non-linear wavenumbers.

Note that all the corrections shown here are less than the cosmic variance uncertainty on $C_l$, even when this cosmic variance is itself conditioned on the local density. It is easy to show that at leading order the conditional cosmic variance is just the familiar $\sqrt{2/(2l+1)}$ in units of the fractional difference, which suggests that for a typical $\delta_0(R)/\sigma(R)$ the correction to the angular power spectrum is negligible, even for close redshift bins. It may be the case that there is a choice of $R$ for which $\delta_0(R)$ is sufficiently large to produce observable corrections. Although Figure~\ref{fig:2Mpp_density} casts doubt on this, a large observable correction is certainly not ruled out.

\begin{figure}
  \includegraphics[width=\columnwidth]{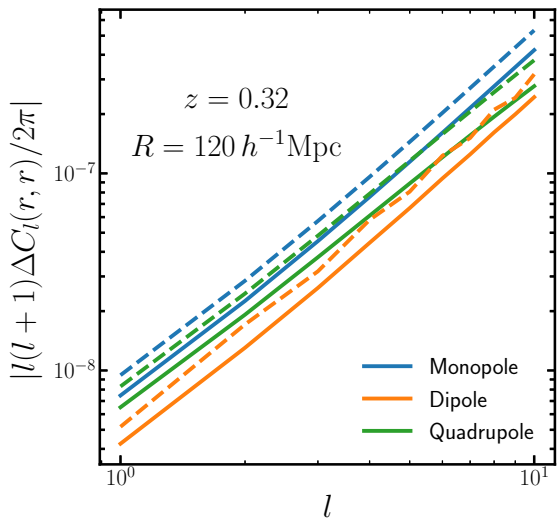}
  \caption{The contributions to the conditional $l>0$ angular power spectrum of matter density fluctuations for a smoothing scale  $R = 120 \, h^{-1} \, \mathrm{Mpc}$ at $z=0.32$. The curves are labelled according to their angular dependence in the $F_2$ kernel and are ordered from top to bottom as monopole (blue), dipole (orange), and quadrupole (green). Dashed curves use a matter power spectrum corrected with \textsc{halofit}. The monopole is the dominant term on all angular scales.}
  \label{fig:cond_Cls}
\end{figure}

In Figure~\ref{fig:cond_Cls} we plot the various contributions to Equation~\eqref{eq:bll0_s}, grouped by their angular dependence in the $F_2$ kernel. In detail, the monopole is the term in Equation~\eqref{eq:bll0_s} proportional to $\xi_R$, the dipole is the term involving $\Omega_R$ and $\xi_R'$, and the quadrupole is the term proportional to $\psi_R$. In this figure we consider a single spherical shell at $z=0.32$, the effective redshift of the BOSS LOWZ sample. The monopole term is dominant on these angular scales, implying the radial dependence of the correction is given roughly by $\xi_R(r)$, as suggested by Figure~\ref{fig:diff_cond_Cls} and Figure~\ref{fig:xi_Rs}. Again, changing the matter power spectrum on non-linear scales can change the results at the 10\% level.

To summarize this section, we have computed the angular power spectrum of density fluctuations conditioned on our smoothed local density contrast. For large smoothing scales a perturbative treatment is possible, with corrections at the ten-percent level for very nearby redshift shells and large angular scales, rapidly dropping with redshift. The maximal size of the correction is in rough agreement with the order-of-magnitude estimate made in Section~\ref{subsec:order}. All corrections appear to be within cosmic variance for the typical local density contrast estimated from the \citetalias{2015MNRAS.450..317C} density field. However, since we have not considered smoothing the remote density contrast, biased tracers, or redshift-space effects, our results are of limited practical use. For this reason, we now turn to computing the impacts on gravitational lensing statistics, also considered in \citetalias{2019MNRAS.486.5061R}. Since lensing inevitably includes contributions from nearby structure in projection, corrections from conditioning are potentially relatively more important.

\section{Application to gravitational lensing}
\label{sec:lensing}

In this section we will apply Equation~\eqref{eq:bll0_s} to the lensing convergence field in direction $\mathbf{\hat{n}}$ for sources at a distance $r$, $\kappa(\mathbf{\hat{n}};r)$. The key quantity to compute is now the mixed bispectrum $b^{\kappa \kappa \delta}_{ll0}(r_1,r_2,0)$ for sources at distance $r_1$ and $r_2$.

Before we proceed we should note that all second-order contributions to the convergence should be included, not just those due to large-scale structure calculated in the previous section. In particular, to be consistent we must include post-Born effects at second-order in $\kappa(\mathbf{\hat{n}})$ -- the relevant cross-bispectrum term at leading post-Born order in the flat-sky approximation has recently been presented in Ref.~\cite{2019JCAP...10..057F}. Post-Born effects are likely to be most relevant for distant sources such as the CMB, where photon trajectories are sufficiently long that the post-Born signal can accumulate and become comparable with large-scale structure terms~\cite{2016JCAP...08..047P, 2018PhRvD..98l3510B, 2019JCAP...10..057F}. For the nearby source redshifts we consider, we expect post-Born terms to be subdominant to second-order density terms, so to keep the analysis simple we will neglect post-Born terms in $\kappa(\mathbf{\hat{n}})$, noting that a consistent second-order analysis could use a full-sky version of the cross-bispectrum from Ref.~\cite{2019JCAP...10..057F} as a starting point.

Neglecting post-Born effects and the subdominant radial derivative terms, the lensing convergence from a source at distance $r_1$ is
\begin{equation}
  \kappa(\mathbf{\hat{n}};r_1) = \int_0^{r_1} \mathrm{d}r_i W(r_i,r_1)\delta(r_i\mathbf{\hat{n}};r_i),
\end{equation}
where the lensing kernel is $W(r_i,r_1) = \frac{3\Omega_mH_0^2}{2c^2}[1+z(r_i)]r_i(r_1 - r_i)/r_1$ in our spatially flat cosmology, and the second argument of the density contrast is its time dependence on the zeroth-order lightcone.

The conditional lensing power spectrum for $l>0$ is then given by
\begin{equation}
  \tilde{C}_l^{\kappa \kappa}(r_1,r_2) = \int_0^{r_1} \mathrm{d}r_i \, W(r_i,r_1)  \int_0^{r_2} \mathrm{d}r_j \, W(r_j,r_2) \tilde{C}_l(r_i,r_j)
  \label{eq:nolimber}
\end{equation}
where $\tilde{C}_l(r_i,r_j)$ is given in Equation~\eqref{eq:bll0_s}. Note that we still to work to second order in the fields, which means that contributions from small $|\mathbf{r}_i - \mathbf{r}_j|$ need to be sufficiently suppressed in the lensing integrals. This means we need to make sure $r_1$ and $r_2$ are sufficiently large and ensure we don't consider small angular scales where our weak non-Gaussianity and tree-level approximations break down. We will check these criteria are met by re-computing all numerical results with \textsc{halofit} power spectra.

The double integral in Equation~\eqref{eq:nolimber} adds significant computational complexity to the evaluation of $\Delta C^{\kappa \kappa}_l$. Fortunately, we can write down a Limber approximation~\cite{1992ApJ...388..272K} which will prove to be accurate over a wide range of angular scales. Readers uninterested in the derivation can skip to the main results given in Equations~\eqref{eq:limber} and~\eqref{eq:hiLlimber}.

\subsection{Limber approximations}
\label{subsec:limber}

The correction to the angular power spectrum for $l>0$ can be written as the sum of three terms given by
\begin{align}
  \Delta^0_l(r_1,r_2) &= \int_0^{r_1} \mathrm{d}r_i \, W(r_i,r_1)  \int_0^{r_2} \mathrm{d}r_j \, W(r_j,r_2) \nonumber \\
  & \, \, \times \int_0^\infty \frac{k^2\mathrm{d}k}{2\pi^2} f_0(k;r_i,r_j) j_l(kr_i)j_l(kr_j) \nonumber \\
  &+ (r_1 \leftrightarrow r_2),\\
  \Delta^1_l(r_1,r_2) &= \int_0^{r_1} \mathrm{d}r_i \, W(r_i,r_1)  \int_0^{r_2} \mathrm{d}r_j \, W(r_j,r_2) \nonumber \\
  & \, \, \times \int_0^\infty \frac{k^2\mathrm{d}k}{2\pi^2} f_1(k;r_i,r_j) j'_l(kr_i)j_l(kr_j)  \nonumber \\
  & + (r_1 \leftrightarrow r_2),\\
  \Delta^2_l(r_1,r_2) &= \int_0^{r_1} \mathrm{d}r_i \, W(r_i,r_1)  \int_0^{r_2} \mathrm{d}r_j \, W(r_j,r_2) \nonumber \\
  & \, \, \times \int_0^\infty \frac{k^2\mathrm{d}k}{2\pi^2} f_2(k;r_i,r_j) j''_l(kr_i)j_l(kr_j)  \nonumber \\
  &+ (r_1 \leftrightarrow r_2),
\end{align}
where the functions $f_0$, $f_1$, and $f_2$ are smooth functions of their arguments. The Limber approximation can be applied to the first term in the standard way, by assuming that the spherical Bessel functions oscillate many times over the characteristic length scales of $f_0$, which allows $f_0$ to be pulled out of the integral. We can then use the orthogonality of the spherical Bessels to do the $k$ integral, which then allows one of the radial integrals to be done, leaving us with one remaining radial integral over a smooth slowly varying integrand, a good approximation even at fairly low $l$. This gives
\begin{equation}
  \Delta^0_l(r_1,r_2) \approx 2\int_0^{r_{\mathrm{max}}} \mathrm{d}r_i \, \frac{W(r_i,r_1)W(r_i,r_2)}{4\pi r_i^2} f_0(l/r_i; r_i, r_i),
  \label{eq:D0_limber}
\end{equation}
where $r_{\mathrm{max}} = \mathrm{min}(r_1,r_2)$. However, this procedure fails with the other two terms, since these involve derivatives of spherical Bessel functions. Instead we follow Ref.~\cite{2018PhRvD..97b3504G} and extend the techniques of Ref.~\cite{2008PhRvD..78l3506L} to derivative terms.

We start with the identity
\begin{equation}
  \int_0^\infty \mathrm{d}r \, f(r)J_\nu(kr) \approx k^{-1}f(\nu/k) + \mathcal{O}[k^{-3}f''(\nu/k)],
\end{equation}
and retain only the leading-order term. In the case of $\Delta_l^0$, we apply this approximation first to the $r_j$ integral and assume that $r_2 \gg \nu/k$ so we can set the upper limit to infinity. This will break down when $r_j$ is close to $r_2$, but this should be benign since the integrand is there suppressed by the lensing kernel. Applying the identity again to the $k$-integral then gives Equation~\eqref{eq:D0_limber} with $l+1/2$ instead of $l$ in the right-hand side, and with an upper limit of $r_1$ instead of $\mathrm{min}(r_1,r_2)$. Clearly the latter choice makes more sense, so we put this limit in by hand.

We can follow the same steps for $\Delta^1_l$ and $\Delta^2_l$ after first converting the spherical Bessel derivatives into undifferentiated $j_l$ with mixed $l$ values, i.e. using
\begin{align}
  j_l'(x) &= \frac{l}{2l+1}j_{l-1}(x) - \frac{l+1}{2l+1}j_{l+1}(x) \\
  j_l''(x) &= \frac{(l-1)l}{(2l-1)(2l+1)}j_{l-2} - \frac{2l(l+1) - 1}{(2l-1)(2l+3)}j_l(x) \nonumber \\
  & \, \, \, + \frac{(l+1)(l+2)}{(2l+1)(2l+3)}j_{l+2}(x)
\end{align}
This gives, for $\Delta^1_l$,
\begin{align}
  &\Delta^1_l(r_1,r_2) \approx \int_0^{\mathrm{min}(r_1,r_2)} \mathrm{d}r \, \frac{W(r,r_1)}{4\pi r^2} \nonumber \\
  & \times \left[\frac{2\nu - 1}{4\nu}\sqrt{\frac{\nu-1}{\nu}}W\left(\frac{\nu}{\nu-1}r, r_2\right)f_1\left(\frac{\nu-1}{r}; r, \frac{\nu}{\nu-1}r \right) \right. \nonumber \\
    &\left. - \frac{2\nu + 1}{4\nu}\sqrt{\frac{\nu+1}{\nu}}W\left(\frac{\nu}{\nu+1}r, r_2\right)f_1\left(\frac{\nu+1}{r}; r, \frac{\nu}{\nu+1}r \right) \right] \nonumber \\
  & + (r_1 \leftrightarrow r_2),
\end{align}
where $\nu = l+1/2$.

For $\Delta^2_l$ we have
\begin{align}
  &\Delta^2_l(r_1,r_2) \approx \int_0^{\mathrm{min}(r_1,r_2)} \mathrm{d}r \, \frac{W(r,r_1)}{4\pi r^2} \nonumber \\
  &\times \left[\alpha_\nu W\left(\frac{\nu}{\nu-2}r, r_2\right)f_2\left(\frac{\nu-2}{r}; r, \frac{\nu}{\nu-2}r \right) \right.\nonumber \\
 &\left. - \beta_\nu W\left(r, r_2\right)f_2\left(\frac{\nu}{r}; r, r \right) \right. \nonumber \\
    &\left. +  \gamma_\nu W\left(\frac{\nu}{\nu+2}r, r_2\right)f_2\left(\frac{\nu+2}{r}; r, \frac{\nu}{\nu+2}r \right) \vphantom{\frac12} \right] \nonumber \\
  & + (r_1 \leftrightarrow r_2),
\end{align}
where
\begin{align}
  \alpha_\nu &= \frac{(2\nu - 3)(2\nu-1)(\nu-2)^{\frac{1}{2}}}{16(\nu-1)\nu^{\frac{3}{2}}}, \nonumber \\
  \beta_\nu &= \frac{[(2\nu - 1)(2\nu+1) - 2]}{8(\nu^2-1)}, \nonumber \\
  \gamma_\nu &= \frac{(2\nu +1)(2\nu+3)(\nu+2)^{\frac{1}{2}}}{16\nu^{\frac{3}{2}}(\nu+1)}.
\end{align}
Note that these expressions are only formally valid when $\nu > 2$. 

The correction to the convergence angular power spectrum is, for $l>0$
\begin{equation}
  \Delta C^{\kappa \kappa}_l(r_1,r_2) = \left[\Delta^0_l(r_1,r_2) + \Delta^1_l(r_1,r_2) + \Delta^2_l(r_1,r_2)\right]\frac{\delta_0(R)}{\sigma^2(R)},
  \label{eq:limber}
\end{equation}
with
\begin{align}
  f_0(k;r_i,r_j) &= 4\pi P(k;z_i,z_j)\left[\frac{34}{21}\xi_R(r_i) - \frac{4}{21}\psi_R(r_i)\right] \nonumber \\
  f_1(k;r_i,r_j) &= 4\pi P(k;z_i,z_j)\left[k\Omega_R(r_i) - \frac{\xi'_R(r_i)}{k}\right]  \nonumber \\
  f_2(k;r_i,r_j) &= 4\pi P(k;z_i,z_j)\left[-\frac{4}{7}\psi_R(r_i)\right].
\end{align}

In the limit that $l \gg 1$ we have $\alpha_\nu \rightarrow 1/4$, $\beta_\nu \rightarrow 1/2$, and  $\gamma_\nu \rightarrow 1/4$. Assuming that $P(k)$ varies much more rapidly than the radial dependence of either $W$, $f_1$ or $f_2$, we can Taylor expand the $k$-dependence of $f_1$ and $f_2$ to give, in the $l \gg 1$ limit
\begin{widetext}
\begin{align}
  \Delta C^{\kappa \kappa}_l(r_1, r_2) &\approx 2\frac{\delta_0(R)}{\sigma^2(R)}\int_0^{\mathrm{min}(r_1,r_2)} \mathrm{d}r \, \frac{W(r,r_1)W(r,r_2)}{r^2} \nonumber \\
  &\quad \times \left \{ \left[\frac{34}{21}\xi_R(r) - \frac{4}{21}\psi_R(r)\right]P(l/r; z, z) + \left[\frac{r}{l}\xi_R'(r) - \frac{l}{r}\Omega_R(r)\right]\frac{P'(l/r; z, z)}{r} - \frac{4}{7}\psi_R(r)\frac{P''(l/r; z, z)}{r^2} \right\}
  \label{eq:hiLlimber}
\end{align}
\end{widetext}
where $P'(k) \equiv \partial P(k)/\partial k$ and $P''(k) \equiv \partial^2 P(k)/\partial k^2$ are derivatives of the matter power spectrum. Equation~\eqref{eq:limber} and Equation~\eqref{eq:hiLlimber} are the second main results of this work and represent the corrections to the lensing convergence angular power spectrum from conditioning on $\delta_0(R)$ in the Limber approximation. In practice the corrections are only relevant for low $l$, so we use the more accurate Equation~\eqref{eq:limber} for numerical work. For all results shown we find the Limber approximation disagrees with the full integration Equation~\eqref{eq:nolimber} to at worst a few percent for $l \geq 10$. For $l \leq 10$ we use the full result of Equation~\eqref{eq:nolimber}.

The correction to the lensing angular power spectrum can be essentially thought of as arising from a multiplicative correction to the linear growth factor $D(r)$ of $1 + \alpha\delta_0(R)\xi_R(r)/\sigma^2(R)$ where $\alpha $ is a numerical factor of order unity. This is only significant for $r$ much less than a correlation length, and since the lensing kernel suppresses the contribution of nearby structure we can expect only a very small correction to $C_l^{\kappa \kappa}$.

\begin{figure*}
  \includegraphics[width=1.5\columnwidth]{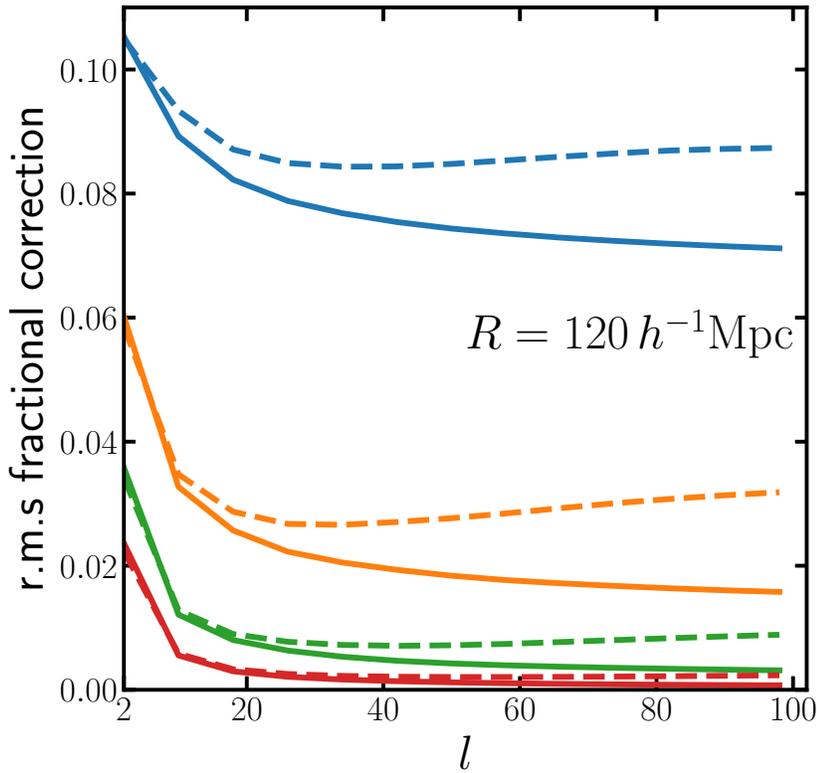}
  \caption{R.m.s. fractional correction to the angular power spectrum of gravitational lensing from a local density fluctuation in a sphere of radius $R = 120 \, h^{-1} \, \mathrm{Mpc}$, for sources at $z_s = 0.1$ (blue top curve), $z_s = 0.2$ (orange second-from-top curve), $z_s = 0.4$ (green second-from-bottom curve) and $z_s = 0.8$ (red bottom curve). Dashed curves use a matter power spectrum corrected with \textsc{halofit}. The correction due to our own local density can be obtained by multiplying these curves by the local density fluctuation $\nu(R) = \delta_0(R)/\sigma(R)$, where our estimate from the \citetalias{2015MNRAS.450..317C} density field is $\nu(120 \, h^{-1} \, \mathrm{Mpc}) \approx 0.85 \pm 0.53$.}
  \label{fig:cond_limber_Cls}
\end{figure*}

In Figure~\ref{fig:cond_limber_Cls} we plot the fractional correction to the lensing angular power spectrum in units of the local fluctuation $\nu(R)$, for a smoothing radius $R = 120 \, h^{-1} \, \mathrm{Mpc}$ and four source redshifts between $z_s = 0.1$ and $z_s = 0.8$. Note that the $l=1$ mode is unobservable in cosmic shear experiments. For reference, the peak of the source distribution for the Euclid lensing survey is expected to be roughly $z_s = 0.8$~\cite{2013LRR....16....6A} and the lower limit of the KiDS+VIKING-450 lensing survey is $z_s = 0.1$~\cite{2018arXiv181206076H}. The figure shows that $\sim 10\%$ corrections to the angular power spectrum can arise at $l=2$ and $z_s = 0.1$ and are at the percent level for higher source redshifts and smaller angular scales. For $l \gtrsim 10$ the corrections are sub-percent for $z_s \gtrsim 0.4$. Note that the sign of the correction is positive on all scales, i.e. conditioning on a positive local density contrast enhances the lensing power. As in the case of the density angular power spectrum, corrections are smaller than the (conditional) cosmic variance on all scales. At the expected peak redshift of the Euclid source distribution the corrections are completely negligible for our estimated value of $\delta_0(R)$.

The dashed curves in Figure~\ref{fig:cond_limber_Cls} have a matter power spectrum which includes a \textsc{halofit} prescription for non-linear scales. For low source redshifts we see sizeable corrections to the predicted $C_l^{\kappa \kappa}$ at $l > 20$. This implies that our perturbative treatment of non-Gaussianity may be breaking down in this regime, so the results should be treated with caution. A more accurate treatment would involve running constrained N-body simulations~\cite{2019MNRAS.486.5061R}. Note that the correction must become large at sufficiently low $R$ and $z_s$, since in this regime nearby structure is highly correlated with our local density. For cosmological surveys however the correction appears to negligible.

Our results are in contrast to those of \citetalias{2019MNRAS.486.5061R}, who claim percent-level effects at all $l$ when $\delta_0 = 0.5$ for a Euclid-like lensing survey. Our results are difficult to compare since \citetalias{2019MNRAS.486.5061R} does not quote a smoothing scale for their local density -- however, it is clear from Figure~\ref{fig:cond_limber_Cls} that at the source redshifts expected to be relevant for Euclid the corrections are sub-percent for all $l > 10$. The discrepancy is likely due to \citetalias{2019MNRAS.486.5061R} taking the Gaussian prediction for $l=0$ and applying it to all $l$ rather than computing the bispectrum, which we have shown is the only contribution at $l>0$.

\begin{figure}
  \includegraphics[width=\columnwidth]{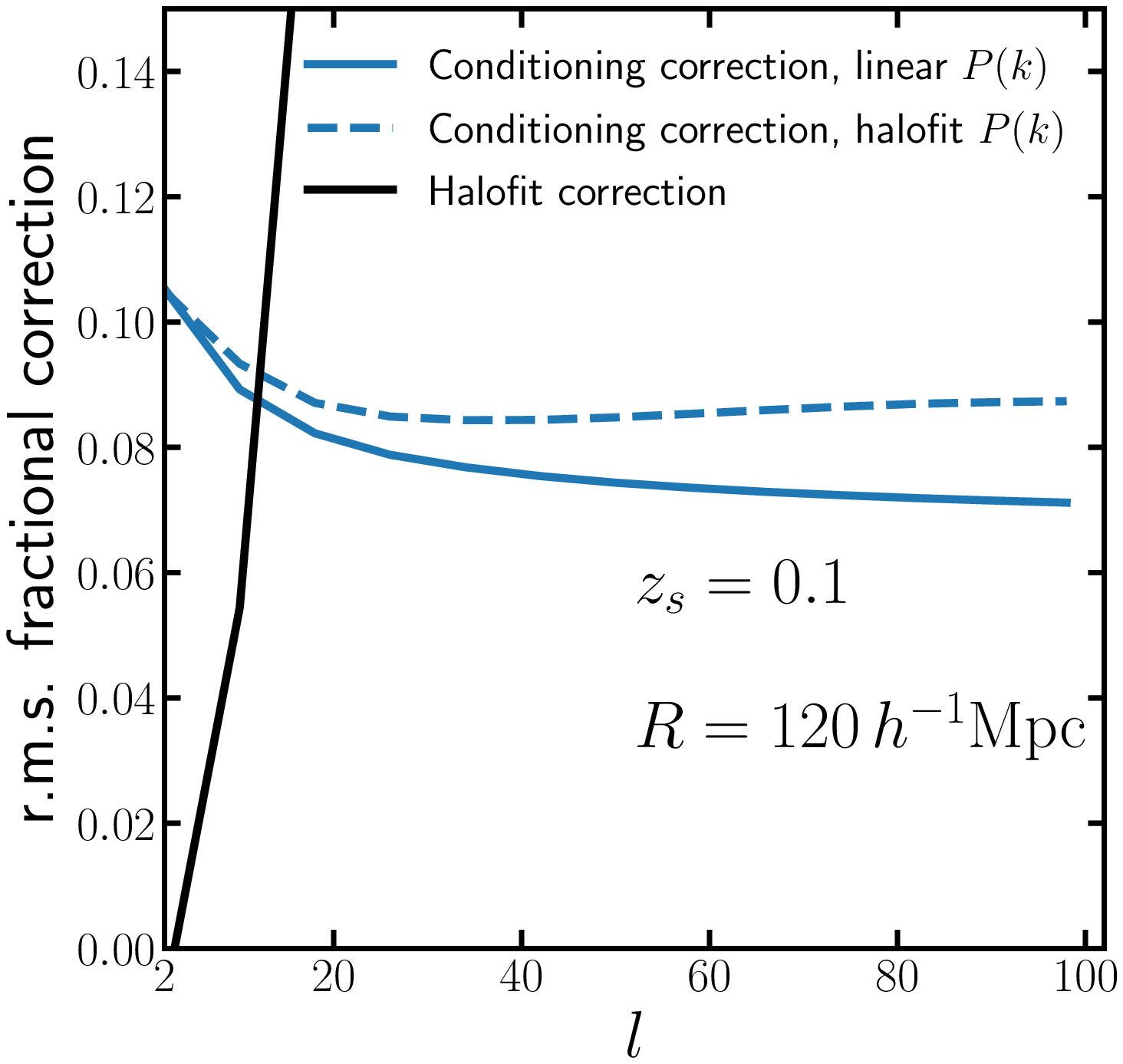}
  \caption{Same as Figure~\ref{fig:cond_limber_Cls} for a source redshift of $z_s = 0.1$ with (blue solid curve) and without (blue dashed curve) a \textsc{halofit} matter power spectrum. We also plot the fractional correction to the lensing angular power spectrum due to non-linearity in the matter power spectrum itself (computed with \textsc{halofit}, black curve). Local corrections to the lensing angular power spectrum can be more important than those due to non-linear clustering of lenses on scales $l \lesssim 10$ and very low source redshifts but are well within cosmic variance for the local density field inferred from 2M++.}
  \label{fig:cl_zs01}
\end{figure}

The largest effects of conditioning on the local density arise at low redshift, where there are large corrections to the unconditional linear lensing power spectrum from non-linear structures. Figure~\ref{fig:cl_zs01} shows the fractional correction to $C_l^{\kappa \kappa}$ for $z_s = 0.1$ in units of the local fluctuation, i.e. the same as the top curve in Figure~\ref{fig:cond_limber_Cls}, along with the fractional correction to $C_l^{\kappa \kappa}$ from non-linearity in the matter power spectrum, estimated by using applying a \textsc{halofit} correction. For the value of $\nu(R)$ estimated in Section~\ref{subsec:order}, $\nu(120 \, h^{-1} \, \mathrm{Mpc}) \approx 0.85 \pm 0.53$, corrections from conditioning are greater than those due to unconditional non-linear structure only at $l < 10$. At the scales most accurately measured by upcoming lensing surveys, corrections due to conditioning are very subdominant to those from non-linear structure for this choice of $R$, and below cosmic variance.

\begin{figure}
  \includegraphics[width=\columnwidth]{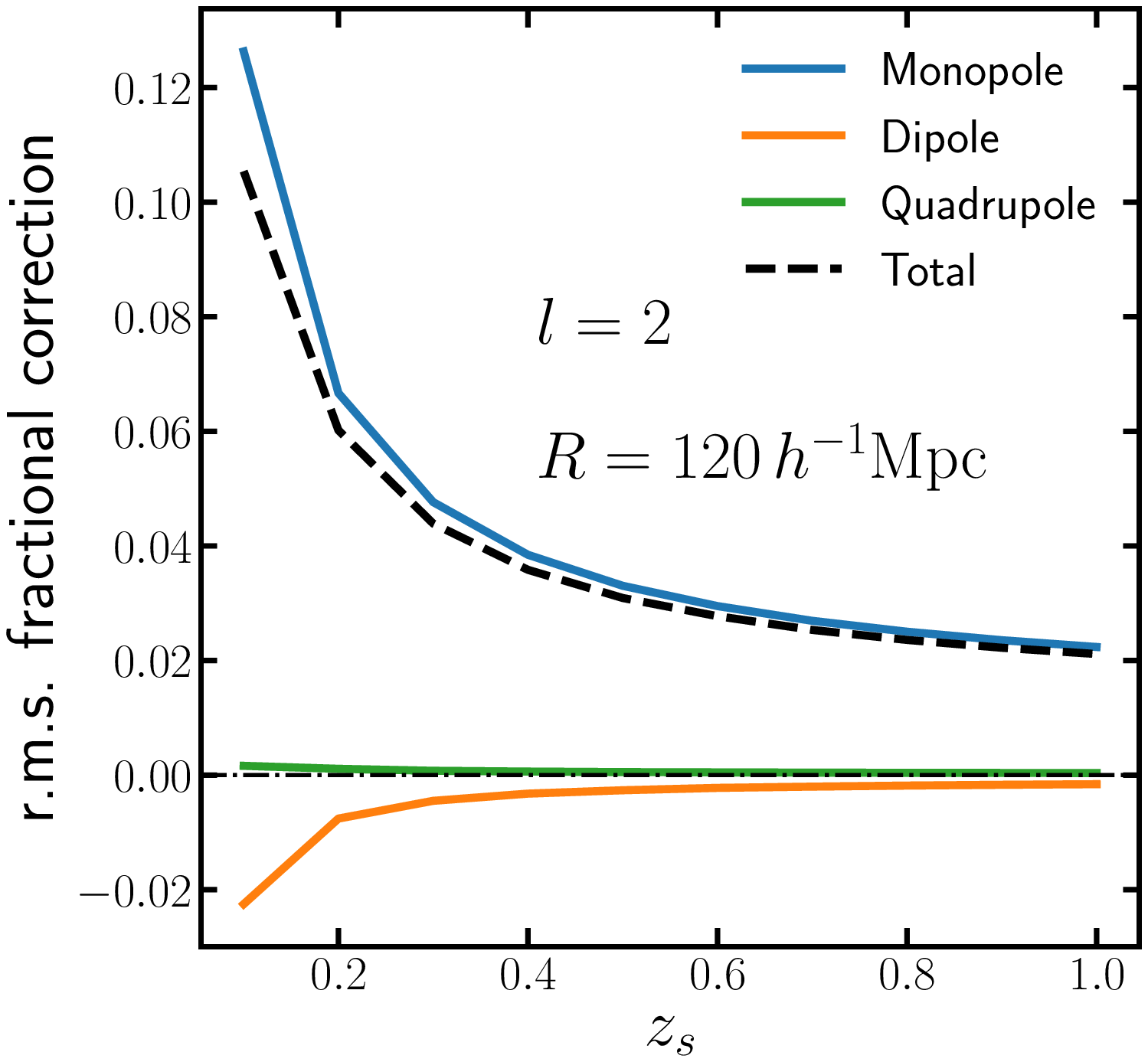}
  \caption{R.m.s. fractional correction to the $l=2$ angular power spectrum of gravitational lensing for sources at redshift $z_s$ and a smoothing scale $R = 120 \, h^{-1} \, \mathrm{Mpc}$. We plot the total correction (black dashed curve) and the individual corrections labelled by their angular dependence in the $F_2$ kernel; monopole (blue upper curve), dipole (orange lower curve) and quadrupole (green middle curve). The dot-dashed horizontal line depicts zero correlation. Corrections can reach $\sim 10\%$ for nearby sources but are still within the conditional cosmic variance. Corrections from non-linear wavenumbers in the matter power spectrum are negligible on these scales.}
  \label{fig:cond_l2_nolimber_Cls}
\end{figure}

In Figure~\ref{fig:cond_l2_nolimber_Cls} we plot the dependence of the correction on source redshift for $l=2$, the angular scale where the correction is largest. As hinted at in Figure~\ref{fig:cond_limber_Cls} the correction drops rapidly with redshift, although the corrections are at the percent level for all values of $z_s$ considered. The dominant contribution is from the monopole term in Equation~\eqref{eq:hiLlimber}, i.e. the term proportional to $P(l/r)$. As suggested by Figure~\ref{fig:NG_corrs} the dominant term is that proportional to $\xi_R(r)$, confirming our earlier claim that conditioning on the local density is roughly equivalent to multiplying the linear growth factor by $1 + \alpha\delta_0(R)\xi_R(r)/\sigma^2(R)$ where $\alpha $ is a numerical factor of order unity, with $\alpha = 34/21$ in Einstein-de Sitter.

\begin{figure}
  \includegraphics[width=\columnwidth]{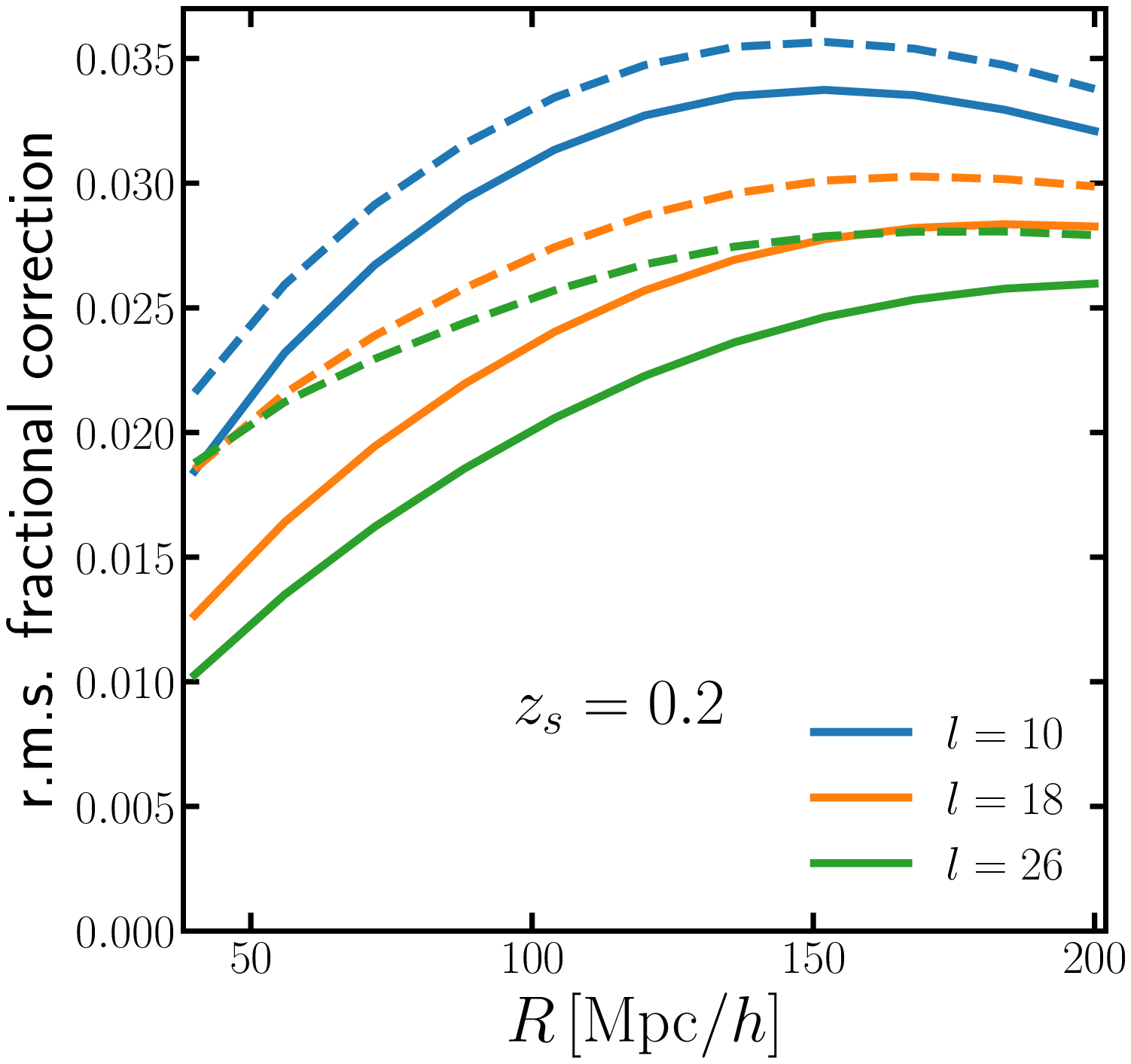}
  \caption{R.m.s. fractional correction to the angular power spectrum of gravitational lensing for sources at redshift $z_s=0.2$ for different smoothing scales $R$ and multipoles $l$. We plot the correction at $l=10$ (blue upper solid), $l=18$ (orange middle solid) and $l=26$ (green lower solid). Dashed curves have been computed using the \textsc{halofit} correction to the matter power spectrum. The dependence on the smoothing scale is modest, although non-linear corrections are likely significant at high $l$ and low $R$.}
  \label{fig:cond_zs0.2_limber_Rs}
\end{figure}

Finally, in Figure~\ref{fig:cond_zs0.2_limber_Rs} we plot the dependence of the r.m.s. correction to lensing angular power on the smoothing scale $R$ for $z_s=0.2$. The dependence is sub-linear at low $R$, with an $l$-dependent turn-over at high $R$. This non-monotonic behaviour is due to the trade-off between the $R$-dependence of the correlation function $\xi_R(r)$ (Figure~\ref{fig:xi_Rs}) and that of the variance $\sigma^2(R)$ (Figure~\ref{fig:2Mpp_density}), since the r.m.s correction from conditioning effectively multiplies the lensing kernel by $\xi_R(r)/\sigma(R)$. While $\sigma(R)$ is a monotonically decreasing function of $R$, $\xi_R(r)$ increases with $R$ at high $r$ and decreases at low $r$, with the relevant values of $r$ depending on $l$ and $z_s$.

If some large fluctuation $\nu^*$ in the local density were found at, say, $R = 200 \, h^{-1} \mathrm{Mpc}$, Figure~\ref{fig:cond_zs0.2_limber_Rs} tells us that the correction from conditioning would be roughly $0.03\nu^*$ with a modest dependence on $l$ on large angular scales and low source redshifts. A 3$\sigma$ fluctuation on this large scale could hence give detectable 10\%-level corrections for the lowest source redshift bins of a Euclid-like survey.

\section{Discussion -- information, bias, and consistency}
\label{sec:discussion}

We have seen that the conditional angular power spectrum of density of lensing fluctuations differs from its unconditional form by a few percent in extreme cases, and at the sub-percent level for the scales and redshifts typically probed by cosmological surveys. In all cases, it appears these differences are smaller than cosmic variance. Nevertheless, it is interesting to ask in what sense neglecting the correction biases measurements of the $C_l$.

The value of our local density contrast $\delta_0(R)$ is extra information which in principle we are free to include or exclude from a cosmological analysis, in the same way as any external data correlated with what we measure in a survey. One might hope therefore that a full joint likelihood analysis with the measured $C_l$ and $\delta_0(R)$ would yield tighter constraints on cosmological parameters than $C_l$ alone. This work has shown that for $l>0$ it is necessary to account for non-Gaussianity in the $C_l$ when computing the correlation, so the exact form of this joint likelihood is complicated. Nonetheless, the correlation between $C_l$ and $\delta_0$ presented in this work is a necessary part of this joint likelihood. Given its small amplitude we do not expect the information gain to be significant.

One could alternatively take the view that since we always observe the Universe from within our local density, neglecting the correction from conditioning amounts to using a biased model for the measured power spectra. For a typical local fluctuation one might naively think that this bias should be smaller than the cosmic variance on the measurement (as we have indeed shown), since the sampling distribution of the measured fields has implicitly marginalized over field values at all spatial locations outside the survey volume, including at the location of the observer. In other words, the variance we assign to our measurement should have implicitly accounted for local density fluctuations, so differences between the unconditional and conditional power spectrum should be less than the cosmic variance (which also includes variance from density fluctuations at all other spatial locations as well as at the observer). This picture is complicated by non-Gaussianity in the sampling distribution of the measurements, but nevertheless one would expect a model conditioned on the local density to provide a better \emph{fit} to the data, since one cannot average out the local fluctuation in the way that typically happens when averaging a summary statistic over the survey volume -- the ergodic theorem will not guarantee convergence of the sample mean statistic to its ensemble mean.

Continuing along these lines, one possible application of the conditional power spectrum is as a consistency check on cosmological statistics. Given a measurement of $\delta_0(R)$, the expressions in this paper may be used to predict conditional angular power spectra. If the cosmological model is consistent, this conditional model should be a good fit to measurements of these power spectra. A similar procedure was used in Ref.~\cite{2019arXiv190712875P} to test for consistency between CMB temperature and polarization measurements in Planck. We have seen that for $\delta_0(R)$ estimated from the 2M++ catalogue and for the scales and redshifts of relevance to cosmological surveys the predicted conditional power spectra are so close to the unconditional power spectra that this consistency test is passed if the unconditional models are good fits to the data. For very low redshift large-scale structure probes the consistency check may be more effective, since the corrections to the unconditional power spectra are expected to be larger.

In principle the expressions derived in this paper could be used to measure the local density field itself on different smoothing scales. In practice we expect only weak upper limits on $|\delta_0(R)|$ will be provided by large cosmology surveys like Euclid. Nevertheless, the possibility of constraining local structure from large-scale structure is an intriguing one.

Finally, pursuing the theme of our local environment and its influence on cosmology one might wonder if the very fact that we exist and are able to make astronomical observations should be information that needs to be conditioned on in a cosmological analysis. When computing ensemble averages of measured summary statistics (such as the lensing power spectrum), we typical allow the underlying density field (or more accurately the primordial curvature fluctuations) to fluctuate at all point in space. If this set of fluctuations includes universes for which the conditions at our spatial location are such that the probability of life forming is very low, should we exclude them from the ensemble? In principle one could test the impact of this with constrained hydrodynamic simulations in the manner of Refs.~\cite{2014MNRAS.438.1805W, 2017MNRAS.467.2787H,2018PhRvD..97j3519H}. One might expect this probability to be related in some way to the local matter density, so the expressions we have derived could be useful in this respect. The uncertainties involved likely preclude a quantitative study, but the concept of `anthropic large-scale structure' is again intriguing.

\section{Conclusions}
\label{sec:conclusions}

The conclusions of this work may be summarised as follows:

\begin{itemize}

\item We have shown that for Gaussian fields, conditioning the angular power spectrum of density fluctuations $C_l$ on the spherically averaged local density $\delta_0(R)$ changes only the $l=0$ power, which is typically unobservable. Changes at $l>0$ arises from non-Gaussianity in the density field, e.g. from non-linear structure formation. The leading order correction is linear in the local density.

\item We have identified a broad range of scales and redshifts where the correction to $C_l$ from conditioning on $\delta_0(R)$ may be computed with a perturbative treatment. As suggested in \citetalias{2019MNRAS.486.5061R} we have used a conditional Edgeworth expansion to show that the leading-order correction is proportional to the bispectrum of the two remote density fields with the local density field. Use of the tree-level bispectrum from second-order perturbation theory and ensuring the local smoothing scale is sufficiently large makes the problem tractable, leading to analytic expressions for the conditional correlation function Equation~\eqref{eq:cond_cov_tree} and the conditional angular power spectrum Equation~\eqref{eq:bll0_s}. These expressions agree with rough analytic expectations in the squeezed limit.

\item For a large smoothing radius $R = 120 \, h^{-1}\mathrm{Mpc}$ chosen to ensure the validity of our perturbative approach, the correction to $C_l$ from a typical value of $\delta_0(R)$ is sub-percent at the redshifts of the BOSS LOWZ galaxy sample. Corrections can reach the percent-level and even ten-percent level for redshifts much smaller than $R$ on large angular scales but are always less than cosmic variance.

\item We have made a rough estimate of $\delta_0(R)$ from the 2M++ density field of \citetalias{2015MNRAS.450..317C} (Ref.~\cite{2015MNRAS.450..317C}), finding that the fluctuation $\nu(R) = \delta_0(R)/\sigma(R)$ is roughly $0.85 \pm 0.53$ for $R = 120 \, h^{-1}\mathrm{Mpc}$. We also made an alternative estimate from 2M++ using the BORG method of Ref.~\cite{2019A&A...625A..64J}, which gives $\nu(120 \, h^{-1}\mathrm{Mpc}) = -0.45 \pm 0.79$. Corrections to the $l>0$ density power spectrum therefore seem to be negligible for current surveys.

\item We have derived the leading-order correction to the lensing angular power spectrum $C_l^{\kappa \kappa}$ in the Born approximation from conditioning on $\delta_0(R)$ and derived its Limber approximant Equation~\eqref{eq:limber} which is accurate for $l \geq 10$. The correction may be approximately captured by modifying the linear growth factor as $D(a) \rightarrow D(a)[1 + \alpha\xi_R(r)\delta_0(R)/\sigma^2(R)]$, where $a$ is the scale factor at comoving distance $r$ on the past lightcone, $\xi_R(r)$ is the linear correlation between $\delta_0(R)$ and $\delta(r)$, and $\alpha$ is a numerical factor of order unity with $\alpha = 34/21$ in Einstein-de Sitter.

\item For a typical value of $\delta_0(R)$ the correction to $C_l^{\kappa \kappa}$ can be at the percent level for source redshifts at $z_s = 0.1$, but quickly becomes small when $z_s$ is increased. For a Euclid-like survey the corrections are sub-percent for $l>10$, in contrast to the claims of \citetalias{2019MNRAS.486.5061R}. The correction is always smaller than cosmic variance and smaller than other non-linear corrections to $C_l^{\kappa \kappa}$ on all but the largest angular scales.

\item In principle failing to apply a correction from conditioning on our local density may bias measurements of $C_l$ from large-scale structure but the bias should be smaller than cosmic variance, as we have indeed found. We would however expect a better fit to the data from using the conditional model, and we have highlighted its use as a consistency check on cosmological models.

\end{itemize}

An additional outcome of this work are the expressions for the conditional cumulants for weakly non-Gaussian fields given in Equation~\eqref{eq:cond_cums} and Equation~\eqref{eq:cond_3pt}. These may be of broader use in astronomical data analysis and generalize previous expressions in the statistics literature.

Note that in this work we have chosen only to condition on $\delta_0(R)$, the local density contrast averaged in spheres around the observer. How valid are the results of this work if a different smoothing kernel were chosen, or non-local but nearby structure were conditioned upon? Firstly note that our conclusion that only non-Gaussianity in the fields can change the $l>0$ power spectrum remains valid for any smoothing kernel, since only a connected three-point function (zero for Gaussian fields) can couple two remote fields with a local field in the required way -- the same is true if we were to condition on the local gravitational potential or the local tidal field. In contrast, if we were to condition on the density at some non-local but nearby overdensity (e.g. the Virgo cluster or the Shapley concentration), we \emph{would} see a scale-dependent effect at the Gaussian level. We defer the study of this possibility to a future work.

In conclusion, we have found no evidence that residing within an overdense region of the Universe induces biases in cosmological statistics in any significant way. As a caveat however we note that we have pursued a perturbative approach to the problem, necessarily limiting the range of scales which can be accurately modelled. A more complete answer to this question will require the running of constrained N-body and possibly hydrodynamic simulations.

\section*{Acknowledgements}
I would like to thank the anonymous referee for useful suggestions which improved the manuscript. I thank Guilhem Lavaux for sharing the BORG data, and Anthony Challinor, Antony Lewis, Geraint Lewis, Robert Reischke, Bj{\"o}rn Malte Sch{\"a}fer, and Blake Sherwin for useful discussions. I acknowledge support from an STFC Consolidated Grant.

\appendix
\section{Reconstruction of the local density field with BORG}
\label{app:borg}

\begin{figure}
  \includegraphics[width=\columnwidth]{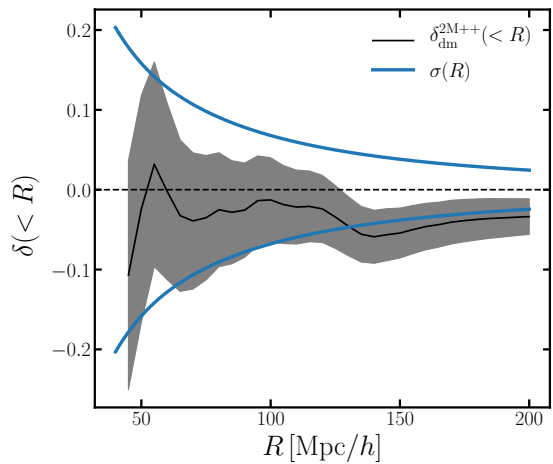}
  \caption{The mean field of the BORG local density field averaged in spheres of radius $R$ around the observer (black solid). The grey bands are estimates of the $1\sigma$ noise on these averages, found by averaging the variance and dividing by the number of voxels in each sphere. The blue solid curves are $\pm$ the linear standard deviation at $z=0$. This figure should be compared with Figure~\ref{fig:2Mpp_density}.}
  \label{fig:borg}
\end{figure}

In this Appendix we present an alternative measurement of the local density field $\delta_0(R)$ using the Bayesian Origin Reconstruction from Galaxies (BORG) method of Ref.~\cite{2019A&A...625A..64J} for reconstructing the dark matter density field. This technique provides samples from the full posterior of the dark matter density field given the 2M++ galaxy catalogue. We take the mean field and average it in spheres of radius $R$ around the observer -- the result is plotted in Figure~\ref{fig:borg}.

We estimate the noise in the spherical average by averaging the variance of each voxel within each sphere and dividing by the number of voxels in that sphere, i.e. we assume that each voxel is independent -- this is an approximation and hence the uncertainties here are lower limits. The noise is about a factor of two larger than our estimate of that of the \citetalias{2015MNRAS.450..317C} field in Section~\ref{subsec:order} but the mean is more robust since more accurate bias models and mean densities have been used. The spherical averages of Figure~\ref{fig:2Mpp_density} are typically no more than $2\sigma$ away from those in Figure~\ref{fig:borg}, in units of the BORG uncertainty, i.e. the two are consistent.

There is a clear preference for lower spherically averaged densities in Figure~\ref{fig:borg} compared with Figure~\ref{fig:2Mpp_density}. At our favoured smoothing radius $R = 120 \, h^{-1}\mathrm{Mpc}$ we find that $\delta_0(R) = -0.024 \pm 0.042$, which is consistent with zero density contrast on this scale. This corresponds to a fluctuation $\nu(R) = -0.45 \pm 0.79$. Setting $\delta_0 = 0$ sets the leading-order correction to the power spectrum to zero as well. We have chosen to present the results of this work in units of the fluctuation $\delta_0(R)/\sigma(R)$ so that any estimate of $\delta_0(R)$ can be inserted. The preference of BORG for density contrasts closer to zero suggests that the corrections to the power spectrum from conditioning may actually be smaller than the r.m.s. values we have found in this work, i.e. safely negligible in future surveys.


\end{document}